\documentclass[aps,prd,preprint,tightening,showpacs]{revtex4-1}
\usepackage{amsmath,amssymb,amsthm,graphicx}
\usepackage[mathscr]{eucal}
\usepackage[colorlinks,linkcolor=blue,anchorcolor=blue,citecolor=green]{hyperref}
\usepackage[dvipsnames,usenames]{color}
\usepackage{graphicx}
\usepackage{subfigure}
\usepackage{amsmath}

\newcommand{\be}{\begin{equation}}
\newcommand{\ee}{\end{equation}}
\newcommand{\ba}{\begin{eqnarray}}
\newcommand{\ea}{\end{eqnarray}}
\usepackage{epsf,epsfig,graphics}
\usepackage{verbatim,color,ulem}
\usepackage{physics}
\usepackage{orcidlink}
\begin{document}
\title{Dynamics of spinning particles in Reissner-Nordstr\"om black hole exterior}
\author{Siang-Yao Ciou \orcidlink{0009-0005-2977-2148}}
\author{Tien Hsieh \orcidlink{0000-0001-7199-1241}}
\author{Da-Shin Lee \orcidlink{0000-0003-3187-8863}}
\email{dslee@gms.ndhu.edu.tw}
\affiliation{Department of Physics, National Dong Hwa University, Hualien, Taiwan, Republic of China}
\date{\today}

\begin{abstract}
We study the orbits of a spinning particle in the Reissner-Nordstr\"om black hole exterior through the spin-curvature coupling to leading order in its spin.
The dynamics is governed by the Mathisson-Papapetrou equations in the pole-dipole approximation.
The equations of motion can be derived and show in particular that in the polar coordinate, the orbits can be restricted to the plane for an aligned spin with the orbital motion, but there is an induced motion out of the plane for a misaligned spin.
The radial potential can be defined from the equation of motion along the radial direction, where the roots are studied to construct the parameter space diagram for different types of orbits.
We then consider the so-called innermost stable circular orbit (ISCO) due to the triple root to see the effects of the particle spin and the black hole charge.
These non-geodesic equations can be solved analytically in terms of the Mino time with the solutions involving Jacobi elliptic functions.
One of the bound motions considered is an oscillating orbit between two turning points in the radial direction.
The usefulness of the solutions is to obtain the periods of the oscillation along the radial direction as well as the induced motion in the polar coordinate for a misaligned spin in both the Mino time and the coordinate time.
Another interesting motions include the inspiral orbit from near ISCO and the homoclinic orbit with the solutions expressed as elementary functions, giving the radial 4-velocity of the inspiral orbit and the Lyapunov exponent associated with the homoclinic orbit.
The implications for gravitational wave emission from extreme mass-ratio inspirals (EMRIs) and black hole accretion are discussed.
\end{abstract}
\maketitle

\section{Introduction}
The first direct observations of gravitational waves from the merger of binary black holes open up a new frontier in
gravitational wave astronomy \cite{abbott-2016, abbott-2019, abbott-2021}.
One of the key sources of low-frequency gravitational waves expected to be observed by the planned space-based Laser Interferometer Space Antenna (LISA) is from extreme mass-ratio inspirals (EMRIs) \cite{consortium-2013, barausse-2020, group-2023, babak-2017, kocsis-2011}.
In astrophysics, EMRIs consist of a stellar mass object orbiting around a massive black hole, and have recently received
considerable attention.

Motivated by EMRIs, the closed-form solutions for the motion of a spinning particle near a spherically symmetric black hole are studied in \cite{witzany-2024} to leading order in a small spin.
These analytical solutions may have applications to the generated gravitational waveforms arising from EMRIs \cite{group-2023, drummond-2022A, drummond-2022B} as well as the understanding of black hole accretion \cite{fabian-2020, page-1974, reynolds-1997, schnittman-2016, jai-akson-2017}.
The motion of a spinless particle around Kerr family black holes along geodesics has been extensively studied in the literature \cite{gralla-2020,fujita-2009} and by us \cite{wang-2022, li-2023, ko-2024}.
The solutions of the trajectories can be written in terms of elliptical integrals and Jacobi elliptic functions \cite{w-1965}, in which the orbits are manifestly real functions of the Mino time \cite{mino-2003} and also the initial conditions can be explicitly specified \cite{gralla-2020}.
In Kerr spacetime, and even for the trajectories of spinning particles, there exist a number of constants of motion in the linear spin order, that makes this particle's dynamics integrable \cite{rudiger-1981, rudiger-1983}. 
Moreover, the separation of the Hamilton-Jacobi equation for the motion of spinning test particles and the corresponding analytical solution have been studied in \cite{witzany-2019, skoupy-2024}.
Our current work considers a spinning particle but orbiting around the exterior of a charged black hole with the Reissner-Nordstr\"om (RN) metric by following \cite{witzany-2024} to explicitly work out the analytical solutions.
For a small spin, we will first derive the radial potential from the underlying equations of motion and classify its roots to construct the parameter space diagram for various types of orbits.
The radius of the innermost stable circular orbits (ISCOs) with the spin effect will be explored \cite{jefremov-2015, zhang-2018}.
Then the closed-form solutions for the motion under consideration will be obtained for both bound and unbound orbits of a particle with an aligned and misaligned spins with respect to its orbital angular momentum.
Theoretical considerations, together with recent observations of structures near Sagittarius A* by the GRAVITY experiment \cite{abuter-2018}, indicate the possible presence of a small electric charge of the central supermassive black hole \cite{zajacek-2018, zajacek-2019}.
Thus, it is of great interest to explore the particle orbiting around a charged black hole.
One of the motions to be studied is an orbit traveling between two turning points along the radial direction, which is a generalization of \cite{witzany-2024} to a charged black hole.
Another interesting orbit is that the particle starts from near ISCO and inspirals into the black hole \cite{mummery-2022, mummery-2023, ko-2024, dyson-2023}.
Lastly, we will also consider the homoclinic orbit, which is the separatrix between bound and plunging orbits and a solution that asymptotes to an energetically bound, unstable spherical orbit \cite{levin-2009, li-2023, stein-2020, ng-2025}.
The existence of the homiclinic orbit is relevant to the chaotic behavior \cite{suzuki-1997} although there is some evidence that terms linear in spin do not cause the emergence of chaos, which in turn has little effect on the overall dynamics of an EMRI with a non-spinning black hole \cite{zelenka-2020}.
The unbound motion starting from space infinity, traveling towards a black hole, hitting the turning point, and returning to space infinity is also shown in Appendix \ref{appendixB} for comparison. 
The Mathematica codes for all calculations are provided by us in \cite{ciou_2025_15233362}.

The article is organized as follows.
In Sec. \ref{sec1}, a review of the equations of motion for a spinning particle in the spherically symmetric RN metric to linear order in spin is provided where their solutions can be recast into integral forms by introducing an effective radial potential.
Section \ref{sec2} focuses on the parameter regime of the particle determined by different roots of the radial potential for both bound and unbound motions.
The radius of the ISCO resulting from the triple root will be studied.
In Sec. \ref{sec3}, we present closed-form solutions of the bound orbits with an aligned and misaligned spins with respect to its orbital angular momentum.
The usefulness of these solutions for the implications for gravitational wave emission from EMRIs and black hole accretion is discussed.
In Sec. \ref{sec4}, the conclusion remarks are drawn.
The roots of the radial potential for a spinless particle are summarized in Appendix \ref{appendixA} \cite{wang-2022}.
Appendix \ref{appendixB} is devoted to the solution of unbound motion.

\section{Equations of motion for a spinning particle}
\label{sec1}
Let us provide a review about the dynamics of a spinning particle orbiting around a black hole \cite{drummond-2022A, witzany-2024}.
A small spinning particle undergoing procession along its trajectory is governed by the Mathisson-Papapetrou equations in the pole-dipole approximation with the degrees of freedom of a monopolar point mass and its spin with the coupling between the curvature and the spin as \cite{corinaldesi-1951, mathisson-2010}
\begin{align}
&\frac{Dp^{\mu}}{d\sigma_m}=-\frac{1}{2} {R^{\mu}}_{\nu\kappa\lambda} \, \dot{x}^{\nu}S^{\kappa\lambda},\label{p_eq}\\
&\frac{DS^{\mu\nu}}{d\sigma_m}=p^{\mu}\dot{x}^{\nu}-p^{\nu}\dot{x}^{\mu},\label{s_eq}
\end{align}
where $D/d\sigma_m$ means a covariant derivative in terms of the proper time $\sigma_m$ along the worldline, and ${R^{\mu}}_{\nu \kappa \lambda}$ is the Riemann curvature of the spacetime.
The particle has 4-momentum $p^{\mu}$ and 4-velocity $\dot{x}^{\nu}$, orbiting around a curved spacetime where the dot is the derivative with respect to the proper time.
In addition, $S^{\mu\nu}$ is the spin tensor of skew symmetry, $S^{\mu\nu}=-S^{\nu\mu}$.
Adopting the Tulczyjew-Dixon spin supplementary condition $S^{\mu\nu}p_{\nu}=0$ in \cite{Tul_1959, dixon-1964} leads to
\begin{align}
p^{\mu}=m\dot{x}^{\mu}+\mathcal{O}\left(s^2\right)
\end{align}
with the mass $m$.
The spin tensor can then be expressed as
\begin{align}
&S^{\mu\nu}=m\epsilon^{\mu\nu\alpha\beta} \, \dot{x}_{\alpha}s_{\beta} \, ,\label{big S}
\end{align}
or equivalently
\begin{align}
&s^{\mu}=-\frac{1}{2} m{\epsilon^{\mu\nu}}_{\alpha\beta}\,\dot{x}_{\nu}S^{\alpha \beta} \label{small S},
\end{align}
where $s^{\mu}$ is the specific spin vector and $\epsilon^{\mu \nu \kappa \lambda}$ is the Levi-Civita pseudo-tensor.
The details can be found in \cite{drummond-2022A, drummond-2022B}.
Here we focus on the dynamics of a spinning particle up to the order $\mathcal{O}{(s^2)}$ \cite{witzany-2024} with the approximate forms of (\ref{p_eq}) and (\ref{s_eq}) as
\begin{align}
&\frac{D^2 x^{\mu}}{d \sigma_m^2}=-\frac{1}{4}{R^{\mu}}_{\nu\gamma\delta} \, {\epsilon^{\gamma\delta}}_{\kappa \lambda} \, \dot{x}^{\nu} \dot{x}^{\kappa} s^{\lambda}\, ,\label{p_eq_s}\,\\
&\frac{D s^{\lambda}}{d\sigma_m}=0\, .\label{s_eq_s}
\end{align}
Notice that the spin $s^\lambda$ is parallel transported along the geodesics.
Later, the existence of the constant of motion can transform the second-order differential equations into first-order ones, which become integrable.
The metric of the Reissner-Nordstr\"om black hole exterior of spherical symmetry can be expressed in the form
\begin{align}
ds^2=-f\left(r\right)dt^2+h\left(r\right)dr^2+r^2\left(d\theta^2+\sin^2\theta d\phi^2\right),
\end{align}
where $f\left(r\right)={1}/{h\left(r\right)}=1-{2M}/{r}+{Q^2}/{r^2}$ with black hole mass $M$ and charge $Q$.
For the case of $Q=0$, the metric is reduced to that of the Schwarzschild black hole.
The Killing vectors of this metric are generators of time translation and rotations around the $x, y, z$ axes
\begin{align}
&\xi^{\mu}_{(t)}=(1,0,0,0),\\
&\xi^{\mu}_{(x)}=(0,0,-\sin\phi,-\cos\phi\cot\theta),\\
&\xi^{\mu}_{(y)}=(0,0,\cos\phi,-\sin\phi\cot\theta),\\
&\xi^{\mu}_{(z)}=(0,0,0,1)\,
\end{align}
with the constant of motion in the form \cite{dixon-1970}
\begin{equation}
C_{(\xi)}\equiv p^{\mu}\xi_{\mu}-\xi_{\rho;\sigma}S^{\rho\sigma}.\label{const}
\end{equation}
To find conserved spin-orbital energy and angular momentum, we write the elements of spin tensor from (\ref{big S}) as
\begin{align}
    & S^{01}=-m\frac{r^{2}\sin\theta}{\sqrt{fh}}\left(\dot{\theta}s^{\phi}-\dot{\phi}s^{\theta}\right)=-S^{10},\\
    & S^{02}=-m\sqrt{\frac{h}{f}}\sin\theta\left(\dot{r}s^{\phi}-\dot{\phi}s^{r}\right)=-S^{20},\\
    & S^{03}=-m\sqrt{\frac{h}{f}}\frac{1}{\sin\theta}\left(\dot{r}s^{\theta}-\dot{\theta}s^{r}\right)=-S^{30},\\
    & S^{12}=m\sqrt{\frac{h}{f}}\sin\theta\left(\dot{t}s^{\phi}-\dot{\phi}s^{t}\right)=-S^{21},\\
    & S^{13}=m\sqrt{\frac{f}{h}}\frac{1}{\sin\theta}\left(\dot{t}s^{\theta}-\dot{\theta}s^{t}\right)=-S^{31},\\
    & S^{23}=m\frac{\sqrt{fh}}{r^2\sin\theta}\left(\dot{t}s^{r}-\dot{r}s^{t}\right)=-S^{32},
\label{Smunu}
\end{align}
and $S^{00}$, $S^{11}$, $S^{22}$, and $S^{33}$ are all zero.
The expressions of the constant of motion for energy $\gamma_m$ and angular momentum $\vec{\lambda}_m$ per unit mass can be written from the above spin tensor and (\ref{const}) as \cite{drummond-2022A, witzany-2024}
\begin{align}
&\gamma_m=-f\dot{t}+\frac{r^2\sin\theta f'(s^{\phi}\dot{\theta}-s^{\theta}\dot{\phi})}{2\sqrt{fh}},\label{e}\\
&\lambda_{mx}=-r^2(\sin\phi\dot{\theta}+\cos\phi\cos\theta\sin\theta\dot{\phi})+\sqrt{\frac{f}{h}}\Big[\sin\theta\cos\phi h(s^{r}\dot{t}-s^{t}\dot{r})\notag \\
&\quad\quad+r\sin\phi\sin\theta(s^{\phi}\dot{t}-s^{t}\dot{\phi})+r\cos\phi\cos\theta(s^{t}\dot{\theta}-s^{\theta}\dot{t})\Big],\label{jx}\\
&\lambda_{my}=r^2(\cos\phi\dot{\theta}-\sin\phi\cos\theta\sin\theta\dot{\phi})+\sqrt{\frac{f}{h}}\Big[\sin\theta\sin\phi h(s^{r}\dot{t}-s^{t}\dot{r})\notag \\
&\quad\quad+r\cos\phi\sin\theta(s^{\phi}\dot{t}-s^{t}\dot{\phi})+r\sin\phi\cos\theta(s^{t}\dot{\theta}-s^{\theta}\dot{t})\Big], \label{jy}\\
&\lambda_{mz}=r^2\sin^2\theta\dot{\phi}+\sqrt{\frac{f}{h}}\Big[\cos\theta h(s^{r}\dot{t}-s^{t}\dot{r})+r\sin\theta(s^{\theta}\dot{t}-s^{t}\dot{\phi})\Big],\label{jz}
\end{align}
where the prime means the derivative with respect to the coordinate $r$.
Now, we introduce an antisymmetric Killing-Yano tensor \cite{tanaka-1996}, given by
\begin{align}
F_{\mu\nu} = r \left( \bar{e}_{\mu}^2 \bar{e}_{\nu}^3 - \bar{e}_{\mu}^3 \bar{e}_{\nu}^2 \right) \label{killing yano tensor}
\end{align}
with the choice of
\begin{align}
\bar{e}_{\mu}^0 &= (\sqrt{f}, 0, 0, 0), \\
\bar{e}_{\mu}^1 &= (0, \sqrt{h}, 0, 0), \\
\bar{e}_{\mu}^2 &= (0, 0, r, 0), \\
\bar{e}_{\mu}^3 &= (0, 0, 0, r \sin{\theta}) ,
\end{align}
obeying
\begin{equation}\label{F_sym}
\nabla_\gamma F_{\alpha\beta}+\nabla_\beta F_{\alpha \gamma}=0 \,.
\end{equation}
We then define a vector from (\ref{killing yano tensor}) to be
\begin{align}
l^\nu = F^{\mu\nu} \dot{x}_\mu
= \left(0, 0, -r \sin{\theta} \, \dot{\phi}, \frac{r}{\sin{\theta}} \, \dot{\theta} \right).
\label{l_4_vec}
\end{align}
This is called the orbital angular momentum 4-vector \cite{drummond-2022A}.
It can be shown that (\ref{l_4_vec}) is parallel transported along the geodesics in general spherically symmetric static spacetimes due to (\ref{F_sym}). As a result, for the spinning particle following the non-geodesics through the curvature and spin interaction, ${D l^{\mu}}/{ d\sigma_m} \propto \mathcal{O}(s)$.
The aligned component of spin with $l^{\mu}$ is an approximate constant of motion to order in linear spin
\begin{align}\label{s_parallel}
&s_{\lVert}\equiv\frac{l^{\mu}s_{\mu}}{\sqrt{l^{\nu}l_{\nu}}}=\frac{l^{\mu}s_{\mu}}{\sqrt{\kappa}} \,,\\
&\frac{ds_{\lVert}}{d\sigma_m}=0+\mathcal{O}(s^2) \,.
\end{align}
Because the angular momentum $\vec{\lambda}_m = ( \lambda_{mx}, \lambda_{my}, \lambda_{mz} )$ remains fixed with a constant direction, based on \cite{witzany-2024}, we can rotate the coordinates into new coordinates $(\theta, \phi) \rightarrow (\vartheta, \varphi)$ where its axis of $\varphi$-rotation points in the direction of $\vec{\lambda}_m$.
With respect to this new coordinate system,
\begin{align} \label{J'}
\lambda_{mx}' = \lambda_{my}' = 0, \quad \lambda_{mz}' = \sqrt{\lambda_{mx}^2 + \lambda_{my}^2 + \lambda_{mz}^2} \equiv \lambda_m \,
\end{align}
in the plane at $\vartheta=\pi/2$ and $\dot \vartheta=0$.
We consider $s=s_\parallel$, where $s^\vartheta = -{s_\parallel}/{r}$ given by (\ref{s_parallel}) with $\kappa=r^4 \sin^2 \vartheta \, \dot\varphi^2$ in the new coordinates and $s_\perp=0$.
Notice that $s \equiv \sqrt{s^\mu s_\mu}=\sqrt{s^2_\perp +s^2_\lVert}$.
Equations (\ref{jx}), (\ref{jy}), and (\ref{jz}) to linear order in $s_\parallel$ can be rewritten in terms of the new coordinates $(t,r,\vartheta, \varphi)$.
Then, together with $p^{\mu} p_{\mu}=-m^2$,
the equations of motion for the 4-momentum $p^\mu$ of the particle can be obtained as \cite{witzany-2024}
\begin{align}
&\frac{r^2}{m} p^r 
=\pm_{r} \sqrt{{{ R}^s_m}(r)} \, ,\label{r_eq}\\
&\frac{r^2}{m} p^t 
=\frac{\gamma_mr^2}{f}+\frac{s_{\lVert}\lambda_mf'r }{2f\sqrt{fh}} \, ,\label{t eq}\\
&\frac{r^2}{m} p^\varphi 
={\lambda_m}+\frac{s_{\lVert}\gamma_m}{\sqrt{fh}} \label{phi eq}
\end{align}
form which one can define the radial potential $R_m^s (r)$,
\begin{align}
{R^s_m}(r) \equiv \frac{r^4}{h} \left( -1 + \frac{\gamma_m^2}{f} - \frac{\lambda_m^2}{r^2} \right) + \frac{s_\parallel \gamma_m \lambda_m (2f - rf') r^2}{(fh)^{3/2}} \,.\label{R}
\end{align}
In general, the radial potential has four roots,
\begin{align}
&R^s_m(r) = (\gamma_m^2 - 1) (r - r_{m1})(r - r_{m2})(r - r_{m3})(r - r_{m4}) \label{R four root}
\end{align}
with the root relation, say $r_{m4} \ge r_{m3} \ge r_{m2} > r_{m1}$ when they all are real-valued. The symbols $\pm_r = {\rm sign} (u^r)$ is defined by the 4-velocity of the spinning particle. The roots of the radial potential in (\ref{R}) for a spinless particle with $s_\parallel=0$ have been studied in \cite{wang-2022}, where the roots are summarized in Appendix \ref{appendixA}.
The roots with the spin correction will be explored later, and can be used to construct the parameter space diagram for different types of orbits.
However, for the misaligned case with nonzero $s_\perp$, giving nonzero $s^t, s^r, s^\varphi$, the condition in (\ref{J'}) leads to the induced $\vartheta$ motion out of the plane at $\vartheta=\pi/2$ given by
\begin{align}
\vartheta(\sigma_m) =\frac{\pi}{2} + \delta\vartheta(\sigma_m)\, ,
\end{align}
where
\begin{align}
\delta\vartheta = \frac{\sqrt{f h} \left( s^r \dot{t} - s^t \dot{r} \right)}{r^2 \dot{\varphi}} + \mathcal{O}(s^2) \,. \label{deltatheta1}
\end{align}
To expand the spin $s^\mu$ in (\ref{s_eq_s}) in terms of the components, we introduce the tetrad $\{ e_{(0)}^\mu, e_{(1)}^\mu, e_{(2)}^\mu, e_{(3)}^\mu \}$ from \cite{witzany-2024}, which are orthogonal to each other and also parallel transported along the geodesics.
We start from the first leg $e_{(0)}^\mu$ to be the 4-velocity of the geodesic with $\dot \vartheta =0 $ and $\vartheta=\pi/2$, given from (\ref{r_eq}), (\ref{t eq}), and (\ref{phi eq}) with $s_{\lVert}=0$, as
\begin{equation}
{e}^{\mu}_{(0)} = \left(\frac{\gamma_m }{f} , \dot r, 0, \frac{\lambda_m}{r^2} \right)\, .
\end{equation}
The second leg is chosen to be the normalized $l^\mu$ in (\ref{l_4_vec}), namely $e_{(3)}^\mu={l^\mu}/{\sqrt{\kappa}}$, where $\kappa=r^4 \sin^2 \vartheta \, \dot\varphi^2$.
With $\dot \vartheta =0 $ and $\vartheta=\pi/2$, the tetrad ${e}^{\mu}_{(3)}$ is given by
\begin{equation}\label{e_3}
{e}^{\mu}_{(3)} = \left(0, 0, -\frac{1}{r}, 0 \right)\, .
\end{equation}
Both are parallel transported and are orthogonal to each other.
Two other vectors, which are orthogonal to the plane of $e_{(0)}^\mu$ and $e_{(3)}^\mu$ and orthogonal to each other, can be constructed by
\begin{align} \label{tetrad}
\tilde{e}^{\mu}_{(1)} &= \left(\frac{\dot r r \sqrt{h}}{\sqrt{f (\lambda_m^2 + r^2)}},\frac{\gamma_m r }{\sqrt{f h (\lambda_m^2 + r^2)}} , 0, 0 \right), \\
\tilde{e}^{\mu}_{(2)} &= \left(\frac{\gamma_m \lambda_m }{f \sqrt{\lambda_m^2 + r^2}}, \frac{\lambda_m \dot{r} }{\sqrt{\lambda_m^2 + r^2}},0, \frac{\sqrt{\lambda_m^2 + r^2} }{r^2} \right)\,
\end{align}
in the orbital rotation plane.
Their linear combinations by introducing the precession angle $\psi$, say ${e}^{\mu}_{(1)}$ and ${e}^{\mu}_{(2)}$,
\begin{align}
{e}^{\mu}_{(1)} &= \cos \psi \, \tilde{e}^{\mu}_{(1)} - \sin \psi \, \tilde{e}^{\mu}_{(2)} \, , \\
{e}^{\mu}_{(2)} &= \sin \psi \, \tilde{e}^{\mu}_{(1)}+\cos \psi \, \tilde{e}^{\mu}_{(2)} \, \label{new_tetrad},
\end{align}
are required to be parallel transported, where the precession angle $\psi(\sigma_m)$ is obtained by integrating
\begin{align} \label{psi eq}
\frac{d\psi}{d\sigma_m} = \frac{\gamma_m \lambda_m}{\lambda_m^2 + r^2}\, ,
\end{align}
which is time-dependent in \cite{van-de-meent-2020}.
In general, the 4-spin $s^\mu$ can be expanded as $ s^\mu= s^{(0)} {e}^{\mu}_{(0)}+ s^{(1)} {e}^{\mu}_{(1)}+s^{(2)} {e}^{\mu}_{(2)}+s^{(3)} {e}^{\mu}_{(3)}$.
The spin supplement condition giving (\ref{small S}) requires $u_\alpha s^\alpha=0$, leading to $s^{(0)}=0$ due to ${e}^{\mu}_{(0)}=u^\mu$.
According to (\ref{s_parallel}) with $e_{(3)}^\mu={l^\mu}/{\sqrt{\kappa}}$ and $s^{(2)} =-{s_{\lVert}}/{r}$.
The spin components $s^{(1)}$ and $s^{(3)}$ are in the orbital rotation plane.
Without losing generality, one can choose $s^{(1)}=0$ and $s^{(3)}=s_\perp$.
Again, with a constant spin magnitude due to (\ref{s_eq_s}), one can write $s_\perp=\sqrt{s^2-s^2_\parallel}$ and the spin vector in a form $s^\mu = s_\parallel {e}^{\mu}_{(2)} +s_\perp {e}^{\mu}_{(3)}$ with the tetrads in (\ref{e_3}) and (\ref{new_tetrad}) as
\begin{align}
s^t &= \sqrt{s^2 - s_\parallel^2} \left( \frac{r \dot{r} \sqrt{h} \sin \psi}{\sqrt{f (\lambda_m^2 + r^2)}} + \frac{\gamma_m \lambda_m \cos \psi}{ f \sqrt{\lambda_m^2 + r^2}} \right),\\
s^r &= \sqrt{s^2 - s_\parallel^2} \left( \frac{\gamma_m r \sin \psi}{\sqrt{f h (\lambda_m^2 + r^2)}} + \frac{\lambda_m \dot{r} \cos \psi}{\sqrt{\lambda_m^2 + r^2}} \right),\\
s^\vartheta &= -\frac{s_\parallel}{r}\, ,\\
s^\varphi &= \sqrt{s^2 - s_\parallel^2} \left( \frac{\sqrt{\lambda_m^2 + r^2} \cos \psi}{r^2} \right).
\end{align}
Then the motion of the slightly off the plane at $\theta={\pi}/{2}$ due to the spin component $s_\perp$ can be studied from the precession angle $\psi$ in (\ref{deltatheta1}) though
\begin{align} \label{precession}
\delta \vartheta = \sqrt{s^2 - s_\parallel^2} \left( \frac{\sqrt{\lambda_m^2 + r^2} \sin \psi}{\lambda_m r} \right).
\end{align}
The analytical solutions for the trajectories of the particle can be obtained in terms of the Mino time $\tau_m$ defined as
\begin{equation}
\frac{d x^\mu}{d \tau_m}={r^2} \frac{d x^\mu}{d \sigma_m}\, .
\end{equation}
We can rewrite the equation of motion as an integral form, given by
\begin{align}
\tau_m(r) - \tau_{mi} = I_{r}, \label{Irintegral}
\end{align}
and
\begin{align} \label{I_r}
I_{r} \equiv \int_{r_i}^r \frac{dr'}{\pm_r \sqrt{R^s_m(r')}}\, .
\end{align}
Similarly, the other variables can also be rewritten into integral form:
\begin{align}
t(r) - t_i &= I_{t} \,,\label{t}\\
\varphi(r) - \varphi_i &= I_{\phi} \,,\label{phi}\\
\psi(r) - \psi_i &= I_{\psi} \,,\label{psi}
\end{align}
where
\begin{align}
I_{t} &\equiv \int_{r_i}^r \frac{dr'}{\pm_r \sqrt{{R}^s_m(r')}} \left( \frac{\gamma_m r'^2}{f} + \frac{s_\parallel \lambda_m f' r'}{2f
 \sqrt{fh}} \right) \,, \label{I0_t}\\
I_{\varphi} &\equiv \int_{r_i}^r \frac{dr'}{\pm_r \sqrt{{R}^s_m(r')}} \left({\lambda_m} + \frac{s_\parallel \gamma_m}{ \sqrt{fh}} \right) \,, \label{I0_phi}\\
I_{\psi} &\equiv \int_{r_i}^r \frac{dr'}{\pm_r \sqrt{{R}^s_m(r')}} \left( \frac{\gamma_m \lambda_m r'^2 }{\lambda_m^2 + r'^2} \right) \, \label{I0_psi}
\end{align}
from (\ref{t eq}), (\ref{phi eq}), and (\ref{psi eq}).
Three functions $I_t$, $I_\varphi$, and $I_\psi$ derived from the integrals (\ref{I0_t}), (\ref{I0_phi}), and (\ref{I0_psi}), respectively, show the general expression as
\begin{small}
\begin{align}
I_{t}(\tau_m) &=
\frac{1}{\sqrt{1-\gamma_m^2}} \Bigg\{ \gamma_m (r_+ + r_-) I_1(\tau_m)
+ \gamma_m I_2(\tau_m)
+ \frac{\gamma_m r_+^4 + s_{\parallel} M\lambda_m \left( r_+ - \frac{Q^2}{s_{\parallel} M} \right)}{r_+ - r_-} I_+(\tau_m) \notag \\
& \quad + \left[ \gamma_m (r_+^3 + r_+^2 r_- + r_+ r_-^2 + r_-^3) - \frac{\gamma_m r_+^4 + s_{\parallel} M\lambda_m \left( r_- - \frac{Q^2}{s_{\parallel} M} \right)}{r_+ - r_-} \right] I_-(\tau_m) \Bigg\} +\gamma_m (r_+^2 + r_+ r_- + r_-^2) \tau_m , \label{I_t} \\
I_{\varphi}(\tau_m) &= \left(\lambda_m + s_{\parallel} \gamma_m\right) \tau_m ,\label{I_phi}\\
I_{\psi}(\tau_m) &= \gamma_m\lambda_m \tau_m - \frac{\gamma_m\lambda_m^2}{\sqrt{1-\gamma_m^2}} \, {\rm Im}\left[ I_{i+}(\tau_m)\right] \label{I_psi}
\end{align}
\end{small}
with
\begin{align}
r_\pm = M \pm \sqrt{M^2 - Q^2},
\end{align}
where the functions $I_1$, $I_2,I_{\pm}$, and $I_{i\pm}$ are defined as
\begin{align}
I_1(\tau_m) &\equiv \sqrt{1-\gamma_m^2} \int_{r_i}^{r(\tau_m)} \frac{r'}{\pm_r \sqrt{R^s_m(r')}}dr' ,\\
I_2(\tau_m) &\equiv \sqrt{1-\gamma_m^2} \int_{r_i}^{r(\tau_m)} \frac{r'^2}{\pm_r\sqrt{R^s_m(r')}} dr' ,\\
I_\pm (\tau_m) &\equiv \sqrt{1-\gamma_m^2} \int_{r_i}^{r(\tau_m)} \frac{dr'}{\pm_r\left(r'-r_{\pm_r} \right)\sqrt{R^s_m(r')}}, \\
I_{i\pm} (\tau_m) &\equiv \sqrt{1-\gamma_m^2} \int_{r_i}^{r(\tau_m)} \frac{dr'}{\pm_r \left(r'\mp i\lambda_m\right) \sqrt{R^s_m(r')}},
\end{align}
to be computed in each case later.\\

\section{Roots of the radial potential and innermost stable circular orbit (ISCO)}
\label{sec2}
The roots of the radial potential in (\ref{R}) for a spinless particle have been studied in \cite{wang-2022}, where the roots are summarized in Appendix \ref{appendixA}.
The roots of (\ref{R}) to first order $s_\parallel$ can be written as
\begin{align}\label{roots}
r_{mi} = r^{(0)}_{mi} + s_\parallel r^{(1)}_{mi} +\mathcal{O}(s_\lVert^2)
\end{align}
with $i=1, 2, 3, 4$, where the zeroth order $r_{mi}^{(0)}$ are shown in Appendix \ref{appendixA}.
The small modifications from the spin $s_\parallel$ are obtained as
\begin{align}
r^{(1)}_{mi} = \frac{-2Q^2 \gamma_m \lambda_m +3M \gamma_m \lambda_m r_{mi}^{(0)} -\gamma_m \lambda_m {r_{mi}^{(0)}}^2}{M \lambda_m^2 -(Q^2 + \lambda_m^2) r_{mi}^{(0)} +3M {r_{mi}^{(0)}}^2 -2(1-\gamma_m^2) {r_{mi}^{(0)}}^3} \,.
\end{align}
With the spin correction, the sum of the roots in (\ref{root_relation}) still applies.
Alternatively, one can define the effective potential of the energy independence \cite{liu-2017} by rewriting $R^s_m(r)$ as
\begin{equation}
\begin{aligned}\label{Rs}
R^s_m(r) &= \left(\alpha \gamma_m^2+\beta \gamma_m+\gamma\right)\\
&= \alpha \left(\gamma_m-\frac{-\beta+\sqrt{\beta^2-4\alpha\gamma}}{2\alpha}\right) \left(\gamma_m-\frac{-\beta-\sqrt{\beta^2-4\alpha\gamma}}{2\alpha}\right),
\end{aligned}
\end{equation}
where
\begin{align}
\alpha&=r^4,\\
\beta&=4\lambda_mQ^2s_{\lVert}-6\lambda_mMrs_{\lVert}+2\lambda_mr^2s_{\lVert},\\
\gamma&={-\lambda_m^2Q^2+2\lambda_m^2Mr-\lambda_m^2r^2-Q^2r^2+2Mr^3-r^4.}
\end{align}
The effective potential of relevance is thus,
\begin{align}
V^s_{\rm eff}=\frac{-\beta+\sqrt{\beta^2-4\alpha\gamma}}{2\alpha} \,. \label{Veff}
\end{align}

\begin{figure}[htp]
\begin{center}
    \includegraphics[width=16cm]{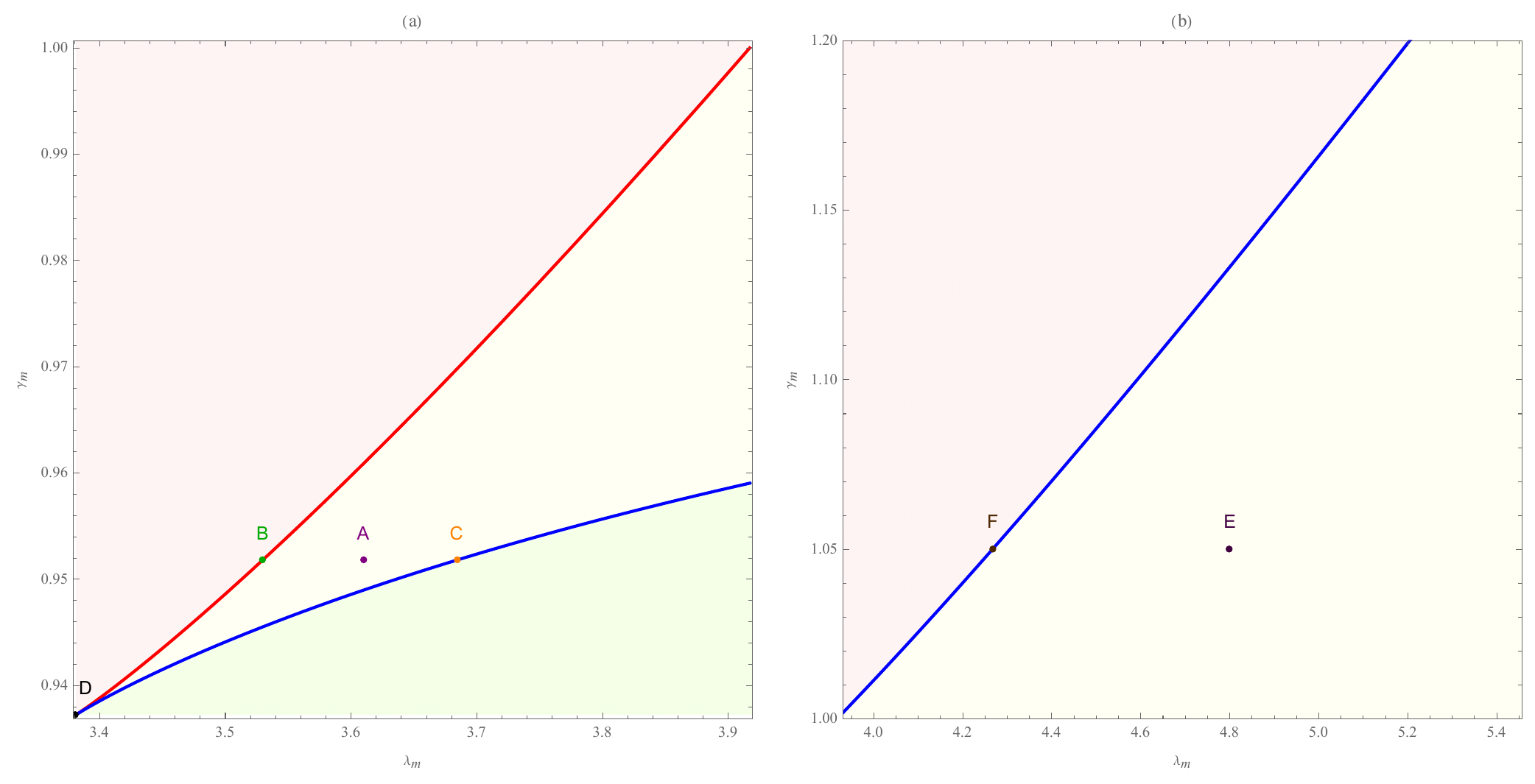}
    \caption{Diagram of parameter space $(\lambda_m,\gamma_m)$ with fixed charge $Q/M=0.5$ and small spin $s_\parallel/m =0.1$ for (a) the bound motion $\gamma_m<1$ in Sec. \ref{sec3} and (b) the unbound motion $\gamma_m>1$ in Appendix \ref{appendixB}.
    The blue line shows the parameters for the stable (unstable) double root of $r_{m3}=r_{m4}$ for bound (unbound) motion and the red line for the unstable double root of $r_{m2}=r_{m3}$ for bound motion in Sec. \ref{sec3C}.
    For bound motion, two double roots merge at the triple root at the point D in Sec. \ref{sec3B}.
    The light yellow region is for the parameters with four distinct real-valued roots in Sec. \ref{sec3A}.
    The light green (red) region is for the parameters with a pair complex-conjugate roots, $r_{m3}=r_{m4}^*$ ($r_{m3}=r_{m2}^*$) and two real-valued roots $r_{m2}>r_{m1}$ ($r_{m4}>r_{m1}$).}
\label{parameter_space}
\end{center}
\end{figure}

\begin{figure}[htp]
\begin{center}
\includegraphics[width=16cm]{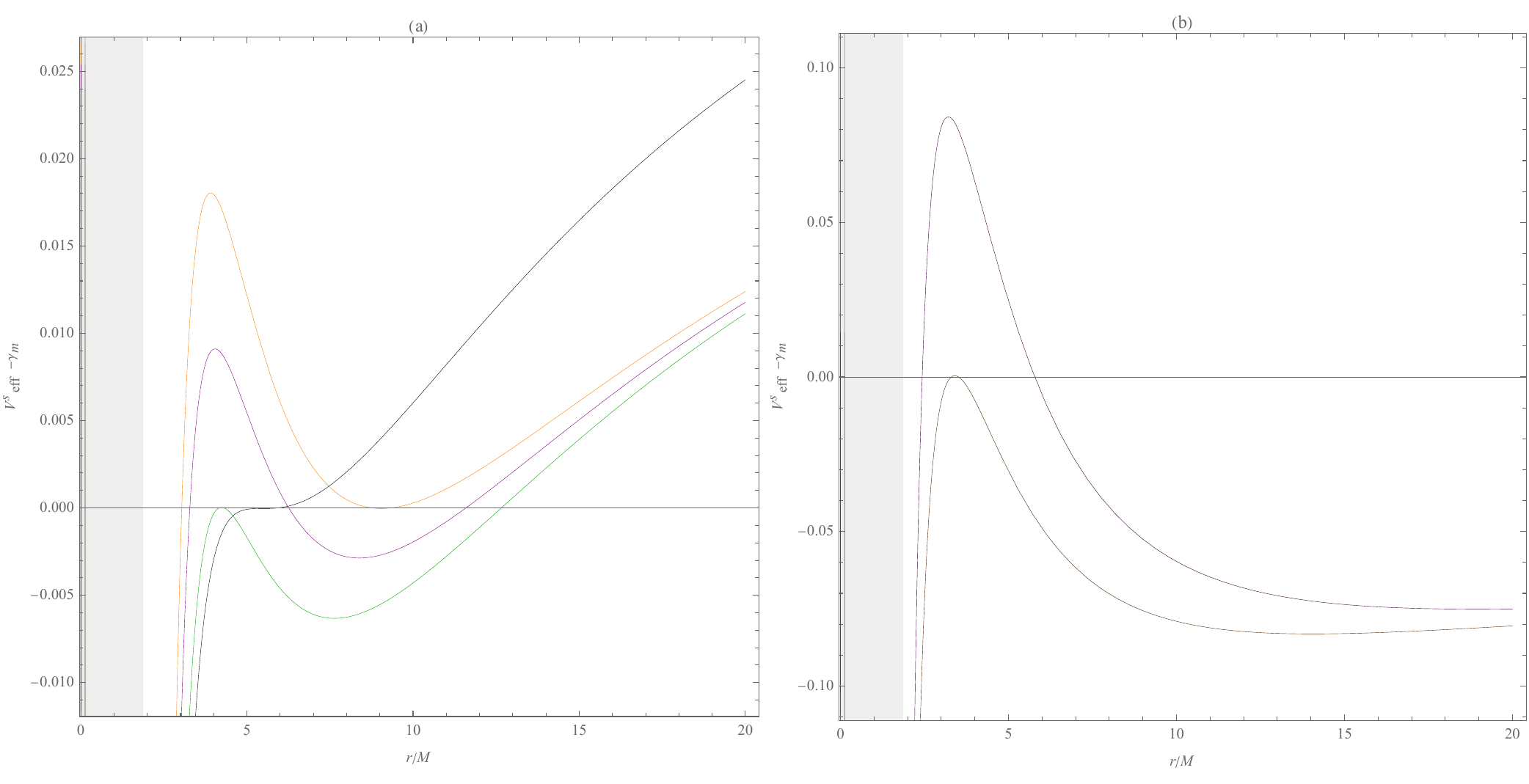}
    \caption{The effective potential $V^s_{\rm eff}(r) -\gamma_m$ as a function of the radial distance $r$ in (\ref{Veff}) with the parameters at points A-D in Fig. \ref{parameter_space}(a) and with the parameters at points E and F in Fig. \ref{parameter_space}(b), where an effective total energy is set to be zero due to (\ref{r_eq}) and (\ref{Rs}).}
\label{veff}
\end{center}
\end{figure}

One of the most interesting orbits is a circular motion with the radius determined by the double root of the radial potential, $R^s_m(r) = 0$ and ${R^s_m}'(r) = 0 $ in (\ref{R}).
In terms of the effective potential, the double root requires that $V^s_{\rm eff}(r)=\gamma_m$ and ${d V^s_{\rm eff}}/{dr}=0$ at $r=r_{\rm mc}$, giving the radius of a circular motion with $\gamma_{\rm mc}$ and $\lambda_{\rm mc}$ satisfying
\begin{align}
\gamma_{\rm mc}^2 &= \frac{\Delta^2 (\Delta - (Mr_{\rm mc} -Q^2))}{r_{\rm mc}^2 (\Delta - (Mr_{\rm mc} - Q^2))^2} - s_\parallel \frac{\Delta \sqrt{Mr_{\rm mc} - Q^2} (M(Q^2 + r_{\rm mc}^2) - 2Q^2 r_{\rm mc})}{r_{\rm mc}^3 (\Delta - (Mr_{\rm mc} - Q^2))^2},\label{Jmc}\\
\lambda_{\rm mc}^2 &= \frac{(\Delta - (Mr_{\rm mc} - Q^2))(Mr_{\rm mc} - Q^2)r_{\rm mc}^2}{(\Delta - (Mr_{\rm mc} - Q^2))^2} + s_\parallel \frac{\Delta \sqrt{Mr_{\rm mc} - Q^2}(2r_{\rm mc}^2 - 9Mr_{\rm mc} + 8Q^2)}{(\Delta - (Mr_{\rm mc} - Q^2))^2}, \label{Emc}
\end{align}
where
\begin{align}
\Delta = r^2 - 2Mr + Q^2.
\end{align}
In the small spin expansion, to linear order in $s_\parallel$ they are {
\begin{align}
\gamma_{\rm mc}&=\gamma^{(0)}_{\rm mc} +s_{\lVert}\gamma^{(1)}_{\rm mc} +\mathcal{O}(s_\lVert^2),\\
\lambda_{\rm mc}&=\lambda^{(0)}_{\rm mc} +s_{\lVert}\lambda^{(1)}_{\rm mc} +\mathcal{O}(s_\lVert^2)\, ,
\end{align}
}
where $\lambda^{(0)}_{\rm mc}$, $\lambda^{(1)}_{\rm mc}$, $\gamma^{(0)}_{\rm mc}$, and $\gamma^{(1)}_{\rm mc}$ can be read off by the straightforward expansion of (\ref{Jmc}) and (\ref{Emc}) to be
\begin{align}
\gamma_{\rm mc}^{(0)} &=\frac{r_{\rm mc} (r_{\rm mc}-2 M)+Q^2}{r_{\rm mc} \sqrt{r_{\rm mc} (r_{\rm mc}-3 M)+2 Q^2}},\label{Emc0}\\
\lambda_{\rm mc}^{(0)} &=\frac{r_{\rm mc} \sqrt{M r_{\rm mc}-Q^2}}{ \sqrt{r_{\rm mc} (r_{\rm mc}-3 M)+2 Q^2}}, \label{Jmc0}
\end{align}
and
\begin{align}
&\gamma_{\rm mc}^{(1)} =\frac{\sqrt{M r_{\rm mc}-Q^2} \left[2 Q^2 r_{\rm mc}-M \left(Q^2+r_{\rm mc}^2\right)\right]}{2 r_{\rm mc}^2 \left( r_{\rm mc}^2 -3 Mr_{\rm mc} +2 Q^2\right)^{3/2}}, \label{Emc1}\\
&\lambda_{\rm mc}^{(1)}=\frac{\left(r_{\rm mc}^2-2 Mr_{\rm mc} +Q^2\right) \left(2 r_{\rm mc}^2 -9 Mr_{\rm mc} +8 Q^2\right)}{2 r_{\rm mc} \left( r_{\rm mc}^2 -3 Mr_{\rm mc} +2 Q^2\right)^{3/2}}. \label{Jmc1}
\end{align}
The parameter $r_{\rm mc}$ for bound motion can be at either the double root $r_{\rm mc} \equiv r^B_{\rm mcu} =r_{m2} =r_{m3}$ for the radius of an unstable circular motion with the values of $\gamma_{\rm mcu}$ and $\lambda_{\rm mcu}$ along the red line in Fig. \ref{parameter_space}(a) or the double root $r_{\rm mc} \equiv r^B_{\rm mcs} =r_{m3} =r_{m4}$ for the radius of a stable circular motion with the values of $\gamma_{\rm mcs}$ and $\lambda_{\rm mcs}$ along the blue line.
Two double roots merge at a triple root with the parameters at point D, the radius of the ISCO to be discussed later.
In particular, the parameters at point A for four distinct real-valued roots, point B for the unstable double root $r^B_{\rm mcu}$, point C for the stable double root $r^B_{\rm mcs}$, and point D for the triple root of the ISCO radius $r_{\rm isco}$ are of interest to find their orbits.
The corresponding effective potentials are plotted in Fig. \ref{veff}.
For unbound motion, the unstable double root of $r_{\rm mc} \equiv r^U_{\rm mcu} =r_{m3} =r_{m4}$ is drawn in the blue line with the parameters, say at F whereas the parameters at point E give four distinct real-valued roots with its effective potential in Fig. \ref{veff}.

%

%

From the radial potential $R_m^s(r)$ with the angular momentum in (\ref{Jmc}) and the energy in (\ref{Emc}), we can find the triple root denoted by $r_{\rm isco}$ which satisfies three conditions simultaneously,
\begin{align} \label{tri_cond}
\begin{cases}
{R_m^s}(r_{\rm isco}) =0 +\mathcal{O}(s_\lVert^2)\\
{R_m^s}'(r_{\rm isco}) =0 +\mathcal{O}(s_\lVert^2)\\
{R_m^s}''(r_{\rm isco}) =0 +\mathcal{O}(s_\lVert^2)
\end{cases}
\end{align}
up to linear order in the spin $s_\lVert$.
Let us assume the ISCO radius can be expanded as
\begin{equation} \label{iscoroot}
r_{\rm isco}= r^{(0)}_{\rm isco} +s_\lVert r^{(1)}_{\rm isco} +\mathcal{O}(s_\lVert^2),
\end{equation}
where the zeroth-order ISCO radius reads
\begin{equation}
\begin{split}
r^{(0)}_{\rm isco} = 2M &+\sqrt[3]{\frac{8 M^4-9 M^2 Q^2+Q^2 \sqrt{5 M^4-9 M^2 Q^2+4 Q^4}+2 Q^4}{M}}\\
&+\frac{4 M^3-3 M Q^2}{\sqrt[3]{M^2 \left(8 M^4-9 M^2 Q^2+Q^2 \sqrt{5 M^4-9 M^2 Q^2+4 Q^4}+2 Q^4\right)}}
\end{split}
\end{equation}
with the correction from the spin,
\begin{small}
\begin{equation}
r^{(1)}_{\rm isco} =\frac{\sqrt{M r^{(0)}_{\rm isco}-Q^2} \left[6 M^2 r^{(0)}_{\rm isco} \left(2 Q^2 +{r^{(0)}_{\rm isco}}^2\right)-M \left(6 Q^4+25 Q^2 {r^{(0)}_{\rm isco}}^2+3 {r^{(0)}_{\rm isco}}^4\right)+8 Q^2 r^{(0)}_{\rm isco} \left(Q^2+{r^{(0)}_{\rm isco}}^2\right)\right]}{r^{(0)}_{\rm isco} \left[18 M^3 {r^{(0)}_{\rm isco}}^2-6 M^2 r^{(0)}_{\rm isco} \left(4 Q^2 +{r^{(0)}_{\rm isco}}^2\right)+M \left(6 Q^4-3 Q^2 {r^{(0)}_{\rm isco}}^2+{r^{(0)}_{\rm isco}}^4\right)+8 Q^4 r^{(0)}_{\rm isco}\right]}
\end{equation}
\end{small}
in terms of $r^{(0)}_{\rm isco}$.
For $Q=0$, the ISCO radius can be reduced to $r_{\rm isco} =6M -2\sqrt{2/3} s_\lVert$ of the Schwarzschild black hole, which is consistent with (68) in \cite{jefremov-2015}.
%
\begin{figure}[h]
\centering
\includegraphics[width=16cm]{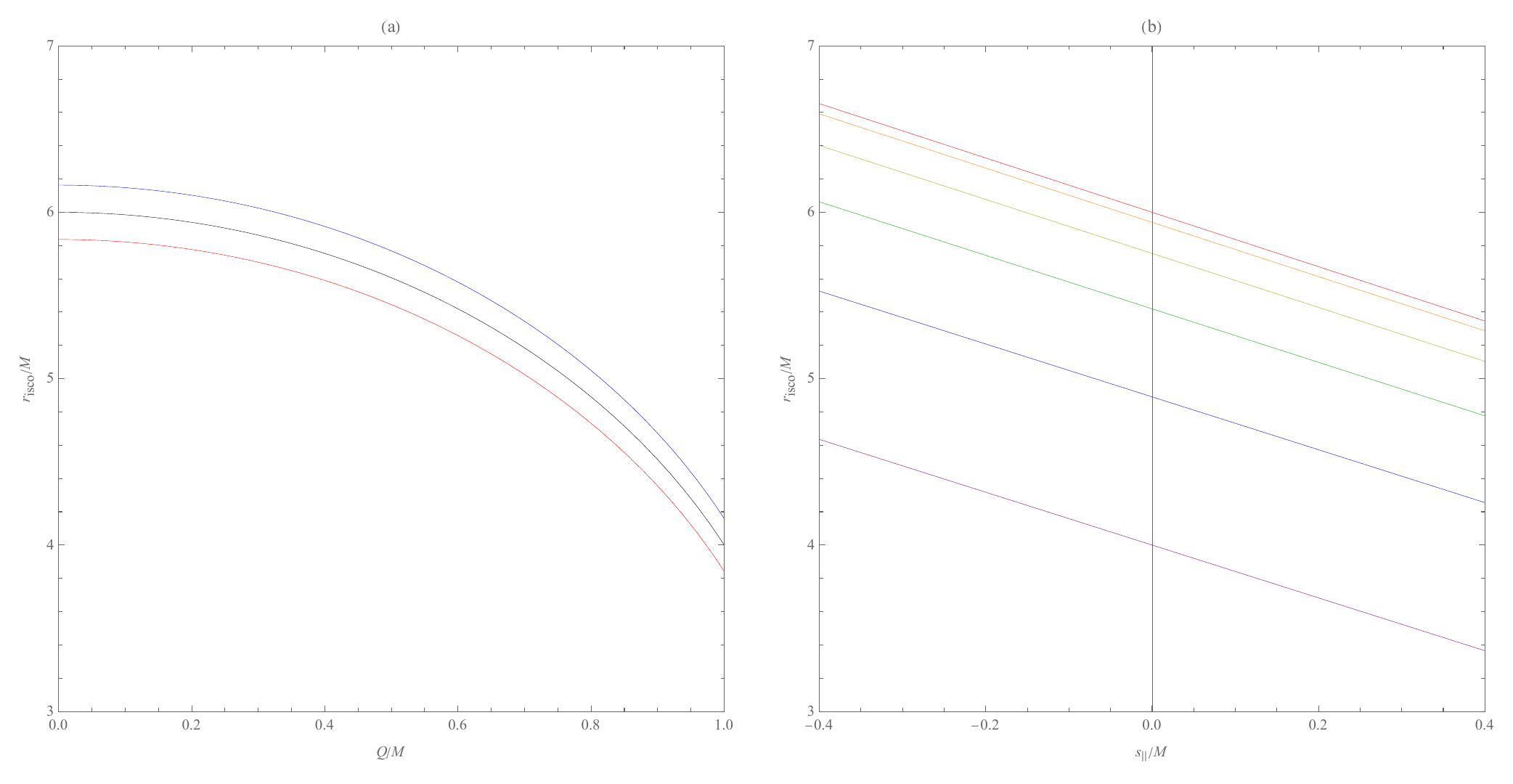}
    \caption{Figures of the radius of the ISCO of the spinning particle as a function of the spin $s_{\lVert}$ and the charge $Q$ of the RN black hole.
    (a):The red (blue) line displays the ISCO radius as a function of $Q/M$ with $\vert s_{\lVert}\vert/m=0.1$ for $s_\parallel >0$ $(s_\parallel <0)$ whereas the black line is for $s_{\lVert}/m=0$;
    (b): the line displays the ISCO radius as a function of $s_\parallel/m$ for the charge $Q/M=0$ (red), $Q/M=0.2$ (orange), $Q/M=0.4$ (yellow), $Q/M=0.6$ (green), $Q/M=0.8$ (blue), and $Q/M=1$ (purple).}
\label{ISCOfig}
\end{figure}
From Fig. 1 in \cite{liu-2017}, it is known that the zeroth-order ISCO radius $r^{(0)}_{\rm isco}$ decreases as the charge increases.
According to Fig. \ref{ISCOfig}, $r^{(1)}_{\rm isco}$ reveals the same behavior.
It is worth mentioning that the spinning particle can have a spin of $s_\lVert >0$ or $s_\lVert <0$, which direction is in the same or opposite with the orbital angular momentum.
For $s_\parallel>0$ ($s_\parallel <0$), the spin effect acts as an effective repulsive (attractive) force, making the ISCO radius smaller (larger) than that for a spinless particle.

\section{Analytical Solutions for Bound Motions}
\label{sec3}
Three interesting bound orbits will be considered below.
With the parameters at point A with four distinct real-valued roots in Fig. \ref{parameter_space}, the spinning particle is trapped between two roots $r_{m4}$ and $r_{m3}$ and oscillates along the $r$ direction between them.
This is an orbit studied in \cite{witzany-2024} but in the Schwarzschild black hole.
Here we will extend its orbit around a charged black hole and also show the consistency with \cite{witzany-2024} in the zero charge limit.
Another interesting orbit has parameters at a triplet root D for the ISCO radius $r_{\rm isco}$.
In this case, we consider the motion toward the black hole with the initial position $r_i$ at $r_i <r_{\rm isco}$.
Thus, the particle will inspiral into the black hole \cite{mummery-2022, mummery-2023, ko-2024}.
The homoclinic orbit will also be obtained with the parameters at point B, starting from the root $r_{m4}$ and moving toward the unstable double root $r_{m3}=r_{m2}$ in \cite{levin-2009,li-2023}.
For unbound motion, the parameters at point E give four distinct real-valued roots, where the orbit can start from the spatial infinity, meet the turning point $r=r_{m4}$ and return to the spatial infinity seen from its effective potential in Fig. \ref{veff}.
Its solution will be presented in Appendix \ref{appendixB} for comparison with those of bound motion.

\subsection{Oscillatory orbit between two turning points}
\label{sec3A}
We start from the bound motion with the parameters of the particle, say at point A in Fig. \ref{parameter_space}. According to the effective potential in Fig. \ref{veff}, the particle will move back and forth along the $r$ direction between two real roots $r_{m3}$ and $r_{m4}$.
Let us set the initial conditions $\tau_{mi}=t_i=0$, $r=r_i$, and $\varphi_i=0$.
To analyze it easily, we rewrite the solutions (\ref{Irintegral}) and (\ref{I_r}) { with a general lower limit of the integral where $r_{m3}\le r_i< r_{m4}$} through the Mino time $\tau_m$ and express it as:
\begin{align}\label{IB0}
{\tau_m(r)}&
=\frac{2\nu_{r_i}}{\sqrt{(1-\gamma_m^2)(r_{m4}-r_{m2})(r_{m3}-r_{m1})}} \nonumber\\
&\times \left\{ F\Bigg[\sin^{-1}\left(\sqrt{\frac{(r_{m4}-r_{m2})(r-r_{m3})}{(r_{m4}-r_{m3})(r-r_{m2})}}\right)\left|{k^{B}}\Bigg]\right.
-F\Bigg[\sin^{-1}\left(\sqrt{\frac{(r_{m4}-r_{m2})(r_i-r_{m3})}{(r_{m4}-r_{m3})(r_i-r_{m2})}}\right) \left|{k^{B}}\Bigg]\right. \right\} \nonumber\\
&\equiv I^B_{r}(\tau_m)
\end{align}
with
\begin{align}\label{kB}
k^{B} &=\frac{(r_{m4}-r_{m3})(r_{m2}-r_{m1})}{(r_{m4}-r_{m2})(r_{m3}-r_{m1})}\, ,\\
\nu_{r_i} &= {{\rm sign}\left(\frac{dr}{d\tau_m}\right) \bigg|_{\tau_m=0}} \,,
\end{align}
and $F(\varphi|k)$ is the incomplete elliptic integral of the first kind.
Using the properties of the Jacobian elliptic function \cite{van-de-meent-2020},
${\rm sn}\left[ F(\sin^{-1}\varphi|k) |k \right] =\varphi$ and
${\rm am}\left[ F(\sin^{-1}\varphi|k) |k \right] =\sin^{-1}\varphi$, the function $\tau_m(r)$ can be inverted to obtain $r(\tau_m)$ as
\begin{align}\label{r_o}
r(\tau_m)=\frac{r_{m3}(r_{m4}-r_{m2})-r_{m2}(r_{m4}-r_{m3}) \,{\rm sn}^2\left(X^{B}(\tau_m)\left|{k^{B}}\right)\right.}{(r_{m4}-r_{m2})-(r_{m4}-r_{m3}) \,{\rm sn}^2\left(X^{B}(\tau_m)\left|{k^{B}}\right)\right.} \,,
\end{align}
where
%
\begin{align} \label{X_tau}
X^{B}(\tau_m)&=\frac{\sqrt{(1-\gamma_m^2)(r_{m4}-r_{m2})(r_{m3}-r_{m1})}}{2}\tau_m
+\nu_{r_i} F\Bigg(\sin^{-1}\left(\sqrt{\frac{(r_{m4}-r_{m2})(r_i-r_{m3})}{(r_{m4}-r_{m3})(r_i-r_{m2})}}\right)\left|{k^{B}}\Bigg)\right.\,
\end{align}
with ${\rm sn}(\varphi|k)$ is the Jacobi elliptic sine function.
The three functions $I_t$, $I_\varphi$, and $I_\psi$ in (\ref{I_t}), (\ref{I_phi}), and (\ref{I_psi}) can be given respectively by
\begin{small}
\begin{align}
I^B_{\pm}(\tau_m)&=\frac{2}{\sqrt{(r_{m3}-r_{m1})(r_{m4}-r_{m2})}} \left[\frac{X^B(\tau_m)}{r_{m2}-r_{\pm}}+\frac{(r_{m2}-r_{m3}) \Pi\left(\beta^B_{\pm};\Upsilon^B_{\tau_m}\left|{k^{B}}\right)\right.}{(r_{m2}-r_{\pm})(r_{m3}-r_{\pm})}\right] -{\mathcal{I}^B_{\pm_i}} \label{IBpm}\;,\\
I^B_{i\pm}(\tau_m)&=\frac{2}{\sqrt{(r_{m3}-r_{m1})(r_{m4}-r_{m2})}} \left[\frac{X^B(\tau_m)}{r_{m2}\mp i\lambda_m}+\frac{(r_{m2}-r_{m3}) \Pi\left(\beta^B_{i\pm};\Upsilon^B_{\tau_m}\left|{k^{B}}\right)\right.}{(r_{m2}\mp i\lambda_m)(r_{m3}\mp i\lambda_m)}\right] -{\mathcal{I}^B_{i\pm_i}} \label{Ii_pm_tau}\;,\\
I^B_{1}(\tau_m)&=\frac{2}{\sqrt{(r_{m3}-r_{m1})(r_{m4}-r_{m2})}} \left[r_{m2}X^B(\tau_m)+(r_{m3}-r_{m2})\Pi\left(\beta^B;\Upsilon^B_{\tau_m}\left|{k^{B}}\right)\right.\right] -{\mathcal{I}^B_{1_i}} \label{IB1}\;,\\
I^B_{2}(\tau_m)&=\nu_{r_i} \frac{\sqrt{\left(r(\tau_m)-r_{m1}\right)\left(r(\tau_m)-r_{m2}\right)\left(r(\tau_m)-r_{m3}\right)\left(r_{m4}-r(\tau_m)\right)}}{\left(r(\tau_m)-r_{m2}\right)} \nonumber\\
&-\frac{r_{m4}\left(r_{m3}-r_{m2}\right)-r_{m2}\left(r_{m3}+r_{m2}\right)}{\sqrt{(r_{m3}-r_{m1})(r_{m4}-r_{m2})}}X^B(\tau_m)-\sqrt{(r_{m3}-r_{m1})(r_{m4}-r_{m2})}E\left(\Upsilon^B_{\tau_m}\left|{k^{B}}\right)\right. \nonumber\\
&+\frac{\left(r_{m3}-r_{m2}\right)\left(r_{m1}+r_{m2}+r_{m3}+r_{m4}\right)}{\sqrt{(r_{m3}-r_{m1})(r_{m4}-r_{m2})}}\Pi\left(\beta^B;\Upsilon^B_{\tau_m}\left|{k^{B}}\right)\right.-{\mathcal{I}^B_{2_i}} \label{IB2}\;.
\end{align}
\end{small}
The parameters of elliptical functions given above follow as
%
\begin{align}
&{\Upsilon^B_{\tau_m}
= {\rm am}\left(X^B(\tau_m) \mid k^{B}\right)
= \nu_{r_i} \sin^{-1} \left( \sqrt{\frac{(r_{m4}-r_{m2})(r(\tau_m)-r_{m3})}{(r_{m4}-r_{m3})(r(\tau_m)-r_{m2})}} \right),} \label{Upsilon} \\
&\beta^B_{\pm} = \frac{(r_{m2}-r_{\pm})(r_{m4}-r_{m3})}{(r_{m3}-r_{\pm})(r_{m4}-r_{m2})}, \\
&\beta^B_{i\pm} = \frac{(r_{m2}\mp i\lambda_m)(r_{m4}-r_{m3})}{(r_{m3}\mp i\lambda_m)(r_{m4}-r_{m2})}, \\
&\beta^B = \frac{r_{m4}-r_{m3}}{r_{m4}-r_{m2}}. \label{beta}
\end{align}
Notice that $\mathcal{I}^B_{\pm_i}$, $\mathcal{I}^B_{i\pm_i}$, $\mathcal{I}^B_{1_i}$, and $\mathcal{I}^B_{2_i}$ depending on the initial conditions $\tau_m=\tau_{mi}=0$ and $r=r_i$ are obtained by requiring ${{I}_{\pm}^{B}(0)={I}_{i\pm}^{B}(0)={I}_{1}^{B}(0)={I}_{2}^{B}(0)=0}$.
In the limit of $Q \rightarrow 0$, the solutions reduce to the expressions in the Schwarzschild case \cite{witzany-2024}.

\begin{figure}[h]
\centering
\includegraphics[width=14cm]{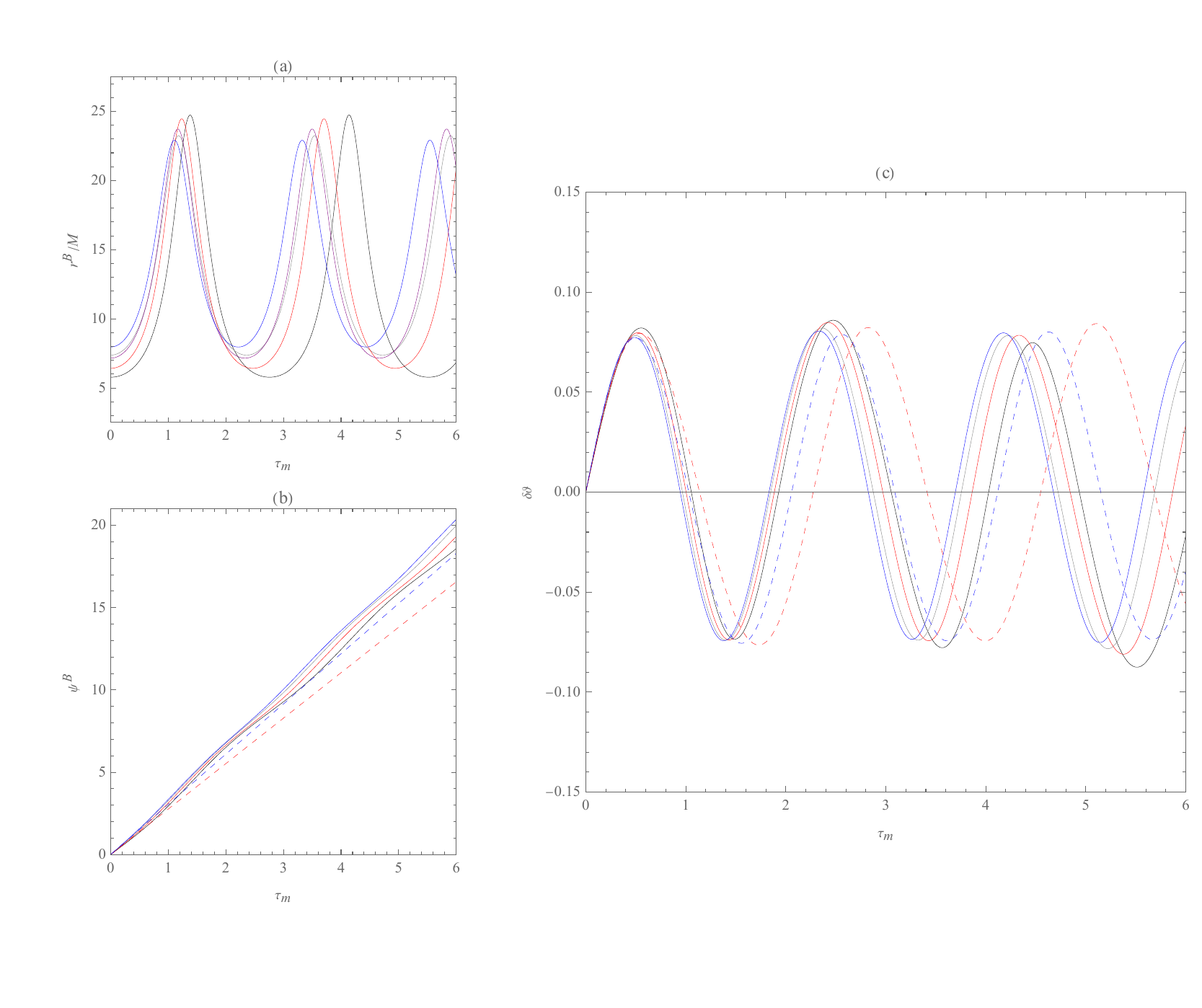}
    \caption{
    Figures of $r^B(\tau_m)$ in (a), $\psi^B(\tau_m)$ in (b), and $\delta \vartheta(\tau_m)$ in (c) of the orbit traveling between turning points as a function of $\tau_m$ for the spinning particle with a misaligned spin,  $s_\perp\neq 0$ and the initial condition of $r$ at $r_i=r_{m3}$.
    The parameters are chosen to be $|s_\parallel|/m= 0.1$, $s/m=0.3$, and $Q/M=0.5$ for $s_\parallel>0$ ($s_\parallel<0$) with red (blue) lines; $|s_\parallel|/m=0.1$, $s/m=0.3$, and $Q/M=0$ for $s_\parallel>0$ ($s_\parallel<0$) with black (gray) lines in the Schwarzschild case; $s/m=0$ and $Q/M=0.5$ for the spinless particle with purple lines.
    The corresponding results given by the approximation $\psi^B(\tau_m) \approx \omega_\vartheta^B \tau_m$ with $\omega_\vartheta^B$ in (\ref{omegaBtheta}) are drawn in dashed line.
    Note that the choice of spin is to emphasize the precession of the polar angle, but in fact, this condition $s_\parallel \gg s_\perp$ should be hold.}
\label{oscillation2}
\end{figure}

\begin{figure}[h]
\centering
\includegraphics[width=14cm]{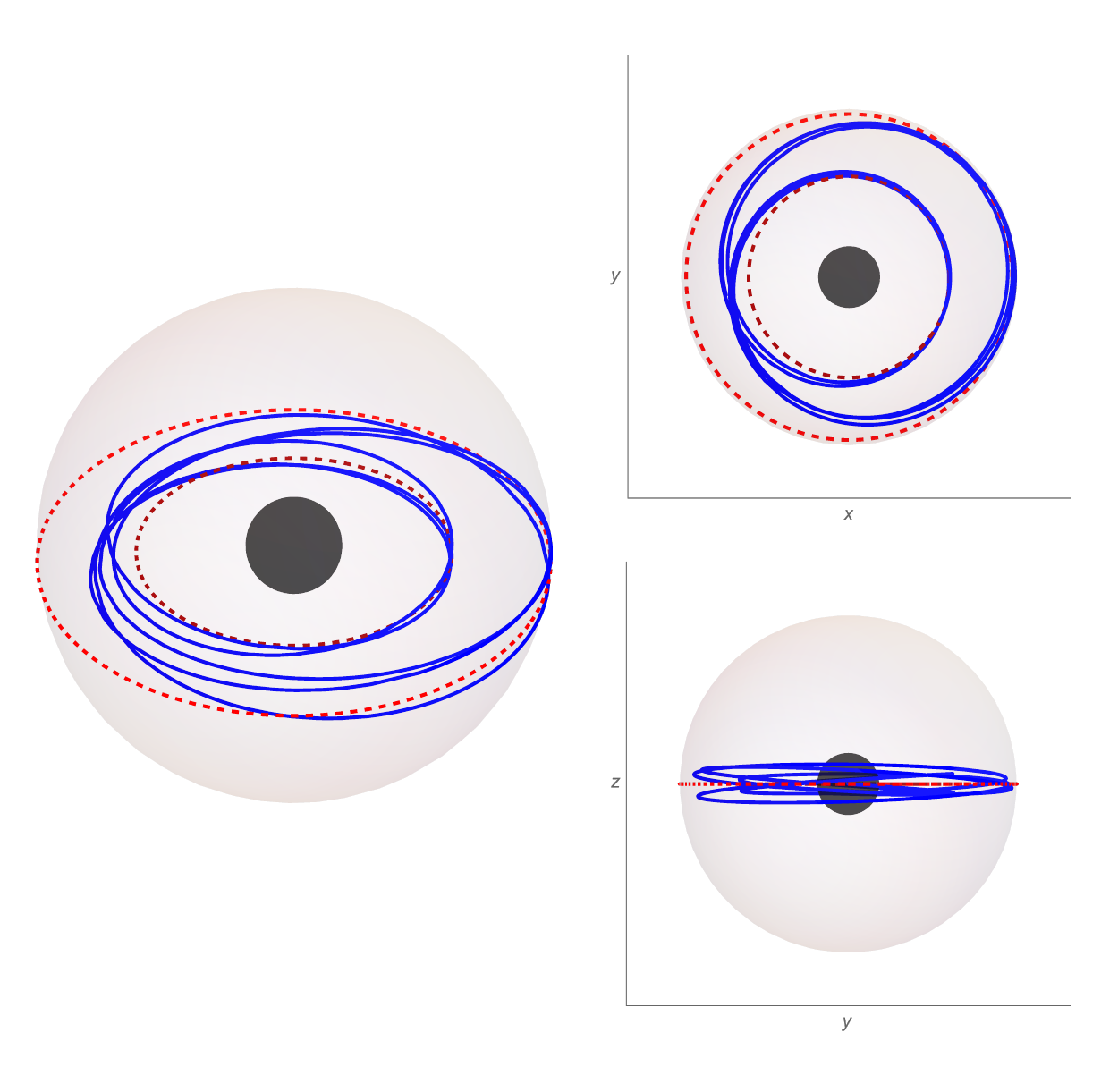}
    \caption{The trajectory traveling between two turning points for the particle with $s_\parallel/m = 0$, $s/m = 0.3$ around the RN black hole with $Q/M=0.5$, which starts from $r_i = r_{m3}$ in the plane at $\theta={\pi}/{2}$ and oscillate between two turning points $r_{m4}$ (the red dashed circle) and $r_{m3}$ (the dark red dashed circle). The induced oscillations of $\delta \vartheta$ is also seen.}
\label{oscillation1}
\end{figure}

The trajectories of a spinning particle along $r$ and $\varphi$ directions for both aligned and misaligned spins are very similar, where Fig. \ref{oscillation2} is for the misaligned case.
The figures show that the particle oscillates between two turning points $r_{m3}$ and $r_{m4}$ with various choices of the particle spin and the black hole charge, which alter not only the turning points, but also the oscillation periods to be obtained later.
In particular, the induced motion along the $\vartheta$ direction for nonzero $s_{\perp}=\sqrt{s^2-s^2_{\parallel}}$ is shown in Fig. \ref{oscillation1}.
To summarize, for a fixed spin $s_\parallel$, the increase in $Q$ gives a relatively larger $r_{m3}$ and smaller $r_{m4}$, leading to a relatively shorter period of oscillation not only along the radial direction but also the induced $\delta \vartheta$ motion than those of the $Q=0$ case.
For a fixed $Q$, the increase in $\vert s_\parallel \vert$ leads to relatively larger (smaller) $r_{m3}$ and smaller (larger) $r_{m4}$ for $s_\parallel <0$ ($s_\parallel >0$) giving relatively shorter (longer) oscillation period along the radial direction as compared with the $s_\parallel =0$ case.
In addition, the period of $\delta\vartheta$ increases from $s_\parallel <0$ to $s_\parallel >0$.
The analytical solutions presented above will show their usefulness in extracting the corresponding oscillation periods in both the Mino time $\tau_m$ and the coordinate time $t$ \cite{fujita-2009, van-de-meent-2020}, which are relevant to the study of the accompanying gravitation wave emission \cite{drummond-2022A}.
The motion of the particle starts from $r_i=r_{m3}$ at $\tau_m(r_{m3})=0$ given by (\ref{IB0}), moves toward $r=r_{m4}$ given by $\tau_m(r_{m4})$, and then return to $r=r_{m3}$ to complete one oscillation with the period
\begin{equation}
{\cal{T}}^B_r= 2 \tau_m(r_{m4}) \,
\end{equation}
in (\ref{IB0}) with $r_i=r_{m3}$.
The oscillation period can be directly read off from the solution of $r(\tau_m)$ in (\ref{r_o}).
The period of ${\rm sn}(X^B,k^B)$ is $4 K(k^B)$, where the quarter-period $K(k)$ is the complete elliptic integral of the first kind \cite{w-1965}.
Therefore the period of ${\rm sn}^2(X^B,k^B)$ is $2 K(k^B)$, which in the Mino time $\tau_m$ becomes
\begin{equation}
{\cal{T}}^B_r= \frac{4 K(k^B)}{\sqrt{(1-\gamma_m^2)(r_{m4}-r_{m2})(r_{m3}-r_{m1})}} \,
\end{equation}
consistent with Fig. \ref{oscillation2}(a).
According to (\ref{t}) and (\ref{I_t}) with $t_i=0$, the period in the coordinate time is given by
\begin{equation}
{{T}}^B_r= 2 t \left(\tau_m= {{\cal{T}}^B_r}/{2}\right)=2 I_t \left(\tau_m= {{\cal{T}}^B_r}/{2}\right) \, ,
\end{equation}
where $I_r^B$, $I^B_\pm$, $I^B_1$, and $I^B_2$ are given by (\ref{IB0}), (\ref{IBpm}), (\ref{IB1}), and (\ref{IB2}) with the vanishing $\mathcal{I}^B_{\pm_i}$, $\mathcal{I}^B_{i\pm_i}$, $\mathcal{I}^B_{1_i}$, and $\mathcal{I}^B_{2_i}$ when $r_i=r_{m3}$ at $t_i=0$.
Moreover, we can obtain the oscillation frequencies $\omega_r^B =2\pi /{\cal{T}}^B_r$ in the Mino time and $\Omega_r^B =2\pi /T^B_r$ in the coordinate time.

The evolution of the precession angle $\psi$ in the early $\tau_m$ in Fig. \ref{oscillation2} can be approximated by $\psi^B(\tau_m) \approx \omega_\vartheta^B \tau_m$, whose slope gives the oscillation frequency of $\delta \vartheta$ given by
\begin{small}
\begin{equation}
\begin{aligned}\label{omegaBtheta}
    \omega_{\vartheta}^B &= \gamma_m \lambda_m +
    \frac{\gamma_m \lambda_m^2}{(\lambda_m^2 + r_{m2}^2) (\lambda_m^2 + r_{m3}^2)}
    \Bigg\{ -\lambda_m \left(\lambda_m^2 + r_{m3}^2\right)\\
    &\quad +\frac{(r_i - r_{m2})^2 (r_{m2} - r_{m3}) (r_{m3} - r_{m4})^2}
    {[ \beta_{i-} (r_i - r_{m3}) (r_{m2} - r_{m4}) - (r_i - r_{m2}) (r_{m3} - r_{m4}) ] [ \beta_{i+} (r_i - r_{m3}) (r_{m2} - r_{m4}) - (r_i - r_{m2}) (r_{m3} - r_{m4}) ] } \\
    &\quad \times
    \left[ -\lambda_m (r_{m2} + r_{m3}) +
    \frac{ (r_i - r_{m3}) (r_{m2} - r_{m4})
    \big[ 2 \lambda_m \, {\rm Re}[\beta_{i+}] (r_{m2} + r_{m3}) +2\, {\rm Im}[\beta_{i+}] (\lambda_m^2 - r_{m2} \, r_{m3}) \big] }
    { 2 (r_i - r_{m2}) (r_{m3} - r_{m4})} \right]\Bigg\} \,.
\end{aligned}
\end{equation}
\end{small}
The oscillation period of $\delta \vartheta$ with respect to the Mino time is thus ${\cal{T}}^B_{\vartheta}= 2\pi/ \omega_{\vartheta}^B$.
Based on the analytical solutions, one can also compute the period with respect to the coordinate time given by
\begin{equation}
{T}^B_{\vartheta} = t(\tau_m=2\pi/\omega_\vartheta^B)
\end{equation}
in (\ref{I_t}) with the obtained functions above.
The periods of oscillation along the coordinates $r$ and $\delta \vartheta$ are some of the main results of this work.
They will provide useful information for studying motion in the frequency domain \cite{drummond-2022B, witzany-2024}.
According to \cite{drummond-2022A, drummond-2022B, witzany-2024}, the analytical solutions, together with the obtained period of oscillation of the orbits, may find their application to couple them to a Teukolsky solver to generate gravitational waves sourced by the motion, thus facilitating the Teukolsky equation computation {\cite{group-2023}}.

%
%

\newpage


\subsection{Homoclinic orbit}
\label{sec3C}
In this section, we consider the homoclinic orbit \cite{levin-2009, li-2023} with the parameters $(\lambda_{\rm mc}, \gamma_{\rm mc})$ at point B in Fig. \ref{parameter_space}.
From the effective potential in Fig. \ref{veff}, the particle starts from the largest root $r_{m4}$, moves toward the black hole, and spends an infinity amount of time to reach the double root $r_{m2}=r_{m3}=r_{\rm mcu}^B$.
The analytical solution for the initial position $r_i=r_{m4}$ can be reduced from those of the oscillation orbit between two turning points for the distinct roots $r_{m4} > r_{m3} > r_{m2}$ in Sec. \ref{sec3A} by exchanging $r_{m3} \leftrightarrow r_{m4}$ and $r_{m1} \leftrightarrow r_{m2}$ for the initial condition which satisfies $r_{m3} < r_i \le r_{m4}$.
The homoclinic solutions have been constructed in Kerr-Newman black holes for a spinless particle in \cite{li-2023}, where part of the solutions with the spin effect share the same form as in \cite{li-2023} but with spin-corrected roots (\ref{roots}) and particle parameters, energy and angular momentum in (\ref{Jmc}) and (\ref{Emc}), in the limit of $a \to 0$ for RN black holes.
Detailed solutions can be referenced in \cite{li-2023} by changing the notation $r_{\rm mss} \to r_{\rm mcu}^B$ and $\nu_{r_i}=-1$.

When $r_{m2} =r_{m3}$ of a double root, the parameter $k^{B}$ in (\ref{kB}) becomes one, causing the elliptic integrals and elliptic functions to reduce to elementary functions (see (27)-(30) in \cite{li-2023}).
Therefore, $\tau_m(r)$ can be obtained
and its inversion $r(\tau_m)$ is given by (31) with $X^{H}(\tau_m)$ of (32) in \cite{li-2023}.
For this case, the integrals $I_t$, $I_\varphi$, and $I_\psi$ are expressed as in (\ref{I_t}), (\ref{I_phi}), and (\ref{I_psi}), where $I_{\pm}^{H}(\tau_m),I_{1}^{H}(\tau_m), I_{2}^{{H}}(\tau_m)$ are given in (35)-(37) in \cite{li-2023}.
The integral of relevance to $\delta \vartheta$ is $I_{i\pm}^{H}(\tau_m)$ expressed as
\begin{small}
\begin{align}
&I_{i\pm}^{H}(\tau_m)=\frac{2}{\sqrt{(r_{\rm mcu}^B-r_{m1})(r_{m4}-r_{\rm mcu}^B)}}\left\lbrace\frac{1}{r_{m1}\mp i\lambda_m}X^{H}(\tau_m)\right.\notag\\
&\left.+
\frac{r_{m1}-r_{m4}}{(r_{m1}\mp i\lambda_m)\left(r_{m4}\mp i\lambda_m\right)\left(h_{i \pm}-1\right)}\left[\sqrt{h_{i \pm}}\tanh^{-1}\left(x^H\left(\tau_m\right)\sqrt{h_{i \pm}}\right)-\tanh^{-1}\left(x^H\left(\tau_m\right)\right)\right]\right\rbrace-\mathcal{I}_{i\pm_i}^{H}\, .\label{I_ipm_H}
\end{align}
\end{small}
In the equations above, we have introduced the notations $x^H (\tau_m)$, $h_{\pm}$, and $h$ defined in (38) and (39) in \cite{li-2023} with an additional one $h_{i\pm}$ below
\begin{align}\label{Upsilon_m_h}
&h_{i\pm}=\frac{(r_{m1}\mp i\lambda_m)(r_{m4}-r_{\rm mcu}^B)}{(r_{m4}\mp i\lambda_m)(r_{m1}-r_{\rm mcu}^B)}\,.
\end{align}
%
In the limit of $r = r_{\rm mcu}^B+\epsilon$ as $\epsilon \to 0$, the time $I^H_{t}$ and the azimuthal angle $I^H_{\varphi}$ have the logarithmic divergence as the double root is approached in \cite{li-2023}.
\\

\begin{figure}[h]
\centering
\includegraphics[width=14cm]{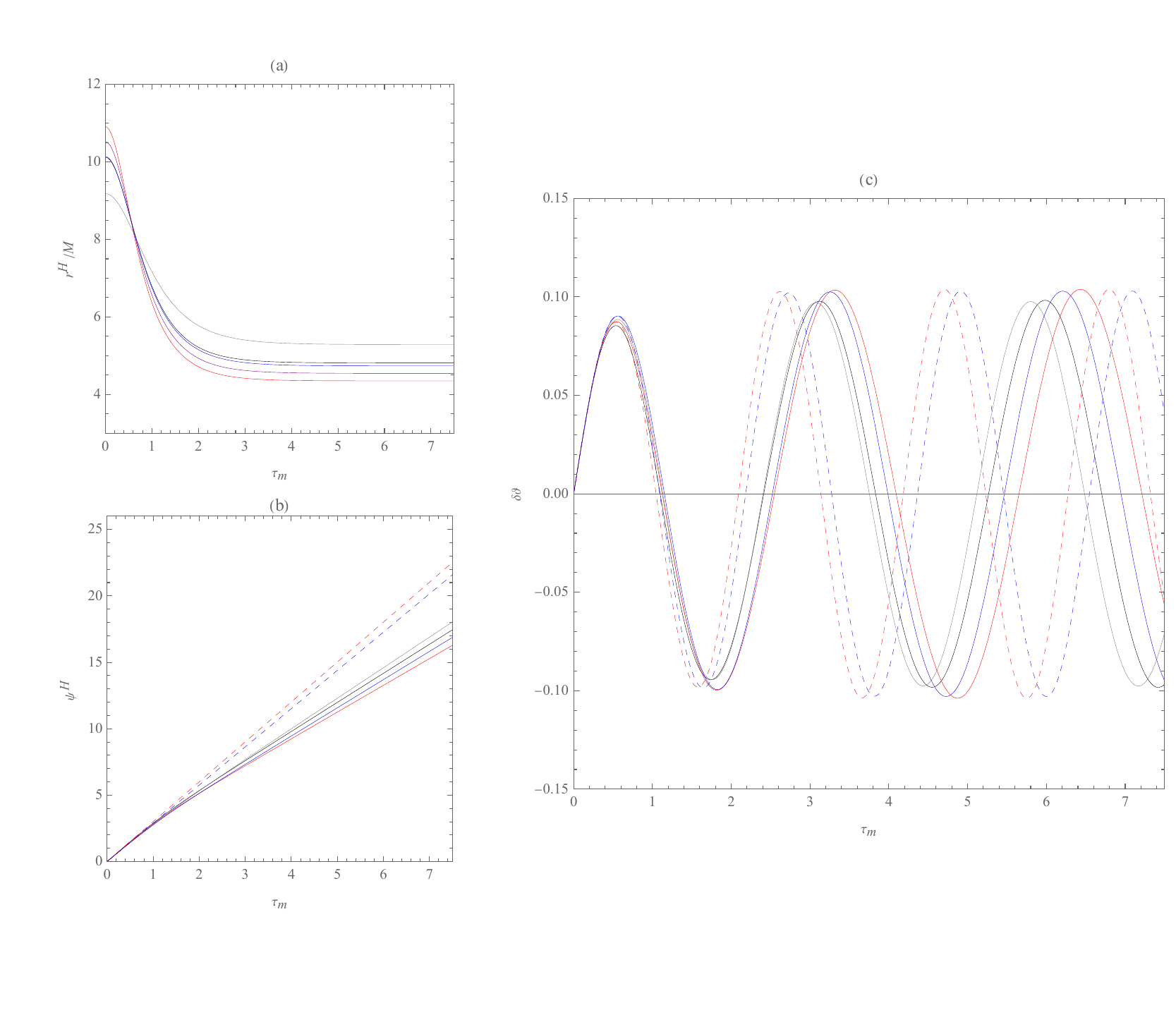}
    \caption{Figures of $r^H(\tau_m)$ in (a), $\psi^H(\tau_m)$ in (b), and $\delta \vartheta(\tau_m)$ in (c) of the homoclinic orbit for the spinning particle with a misaligned spin for the nonzero $s_\perp$ moving around the RN black hole with the charge $Q$.
    The same parameters are chosen as in Fig. \ref{oscillation2} but with the different initial condition in $r$ at $r_i=r_{m4}$ and the choice of the parameters $(\lambda_{\rm mc}, \gamma_{\rm mc})$ from the double root at point B in Fig. \ref{parameter_space}.}
\label{homoclinic2}
\end{figure}

\begin{figure}[h]
\centering
\includegraphics[width=14cm]{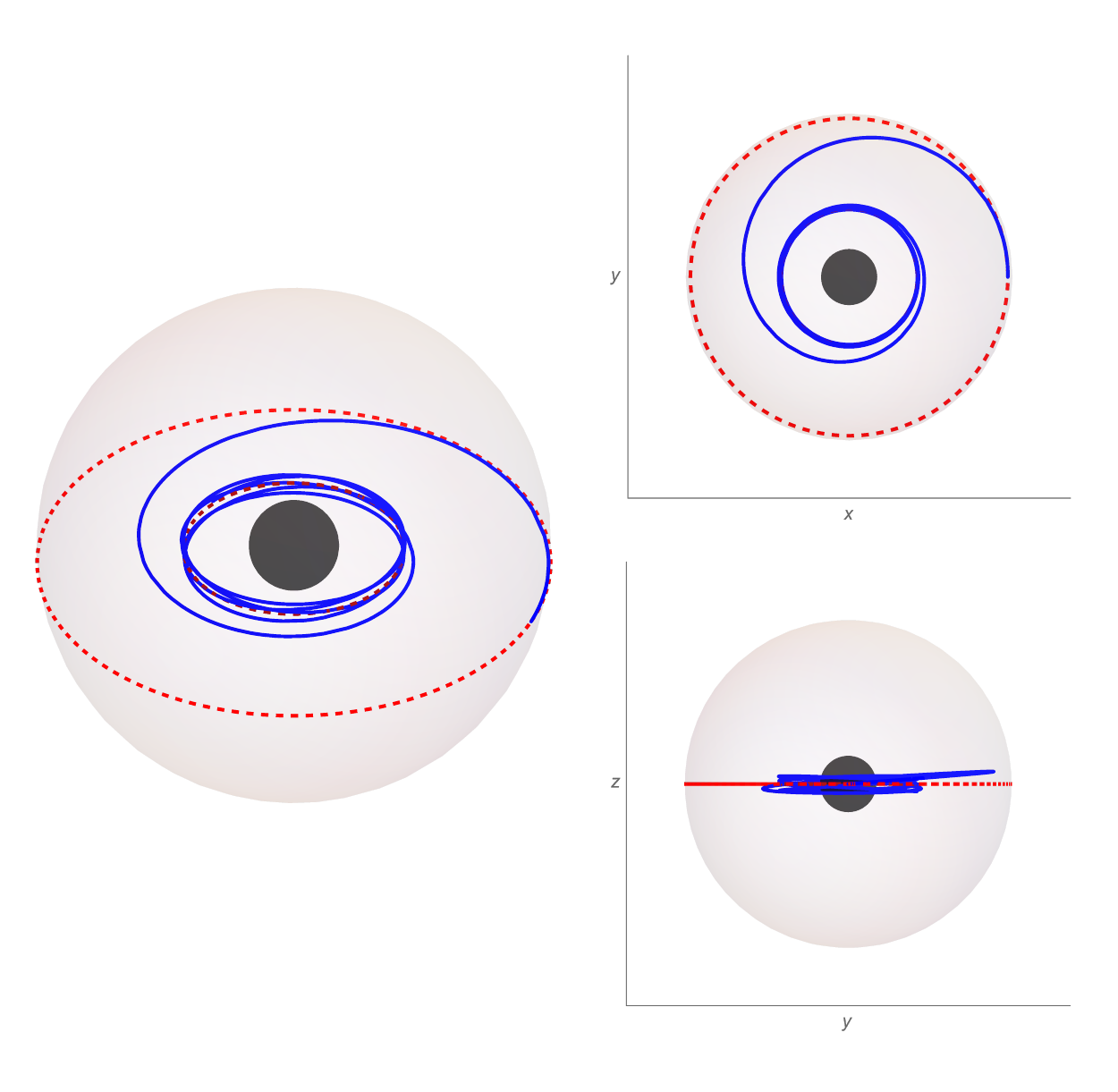}
    \caption{The homoclinic orbit with the same parameters as in Fig. \ref{oscillation2} but with the different initial condition in $r$ at $r_i=r_{m4}$ (the red dashed circle) and the double root $r_{m3}=r_{m2}$ (the dark red dashed circle) with $(\lambda_{\rm mc}, \gamma_{\rm mc})$ at point B in Fig. \ref{parameter_space}.}
\label{homoclinic1}
\end{figure}

The homoclinic orbit for the misaligned spin is plotted in Fig. \ref{homoclinic1}, where the induced oscillatory motion along the $\theta$ direction is seen.
In Fig. \ref{homoclinic2} with various choices of $Q$ and $s$, the precession angle in the early time $\tau_m$ can be approximated by $\psi^H(\tau_m) \approx \omega_\vartheta^H \tau_m$ in (\ref{I_psi}), giving the frequency of oscillation of $\delta \vartheta$ found as
\begin{equation}
\omega_\vartheta^H = \frac{\gamma_m \lambda_m r_i^2}{\lambda_m^2+r_i^2}
\end{equation}
with the period
\begin{equation}
{\cal T}^H_{\vartheta}=2 \pi/ \omega_\vartheta^H
\end{equation}
in terms of the Mino time.
{In Fig. \ref{homoclinic2}, for a fixed value of the spin, as the charge $Q$ increases, the period increases. For a fixed $Q$, the period increases from $s_\parallel <0$ to $s_\parallel >0$.} The period for the coordinate time $t$ can be obtained as
\begin{equation}
{T}^H_{\vartheta}= t(\tau_m=2\pi/\omega_\vartheta^H)
\end{equation}
using the above analytical solutions.
The analytical expression of the period of oscillations of $\delta \vartheta$ is obtained for both the Mino time and the coordinate time given directly from the analytical solution of the orbit. This will be of relevance to the study of the possible chaotic dynamics of a spin particle orbiting around a black hole \cite{suzuki-1997, zelenka-2020}.
\\

%

\begin{figure}[h]
\centering
\includegraphics[width=16cm]{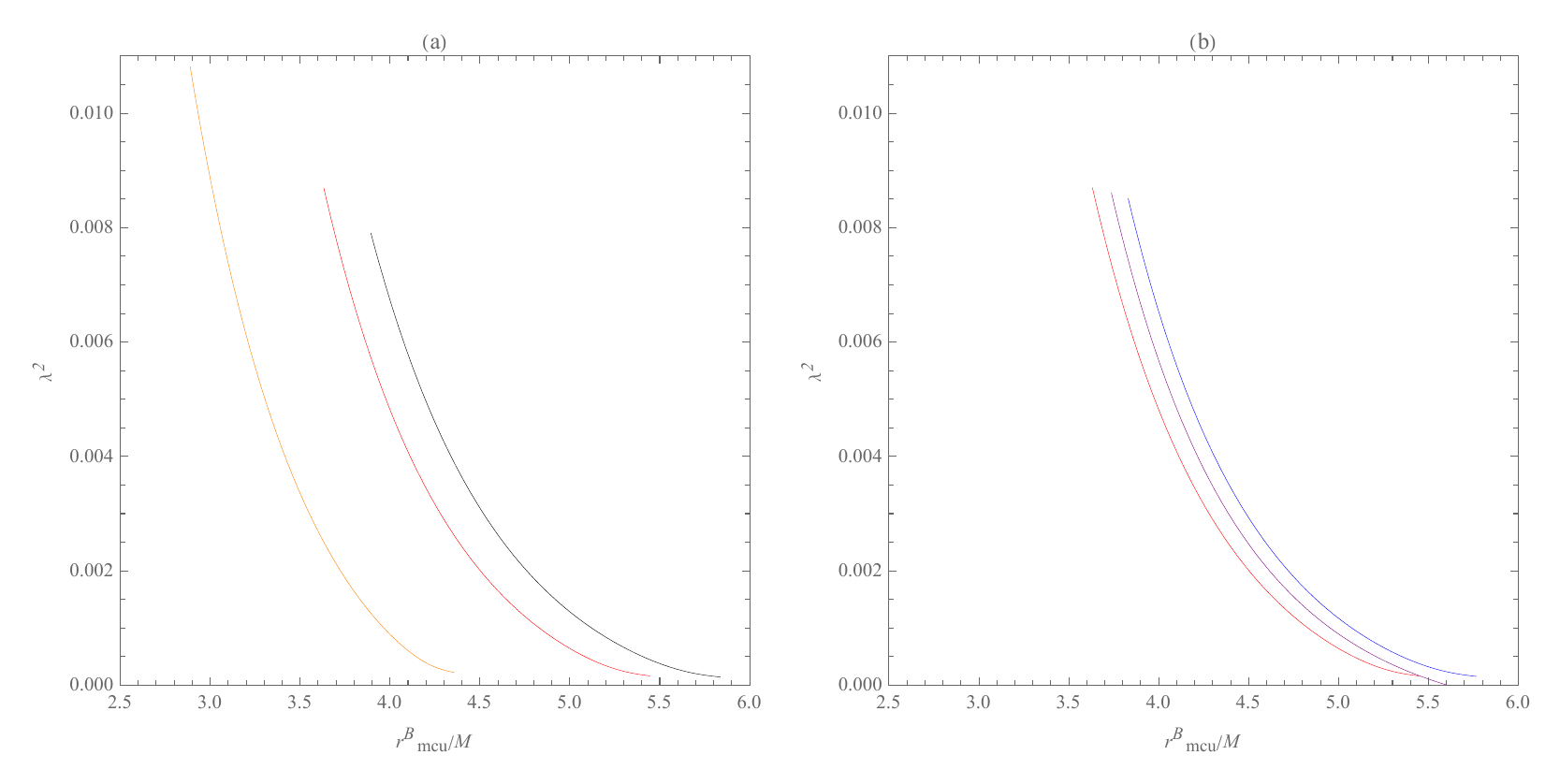}
    \caption{The Lyapunov exponent as a function of $r_{\rm mcu}^B$: (a) for a fixed $s_\parallel/m=0.1$ and $Q/M=0$ (black), $Q/M=0.5$ (red), $Q/M=0.9$ (orange); (b) for a fixed $Q/M=0.5$ and $s_\parallel/m=0$ (purple), $s_\parallel/m=0.1$ (red), $s_\parallel/m=-0.1$ (blue).}
\label{Lyapunov}
\end{figure}

As for the corresponding Lyapunov exponent, according to \cite{cardoso-2009}, it can be determined by the inverse of the instability time scale associated with the orbital motion.
Let us now define the effective potential in the coordinate $t$ from the radial equation of motion
\begin{equation}
\frac{1}{2} \left( \frac{ dr}{dt} \right)^2+\mathcal{V}^s_{\rm eff}(r)=0 \, .
\end{equation}
From (\ref{r_eq}) and (\ref{t eq}),
\begin{equation}
    \mathcal{V}_{\rm eff}^s(r) = -\frac{\left(\gamma_m^2-1\right) (r-r_{m1}) (r-r_{\rm mcu}^B) (r-r_{\rm mcu}^B) (r-r_{m4}) \left( r^2-2 Mr +Q^2\right)^2}{2 \left(\lambda_m M r {s_\parallel}-\lambda_m Q^2 {s_\parallel}+\gamma_m r^4\right)^2}.
\end{equation}
Then, consider the spinning particle near the unstable double root, that is $r(t)=r_{\rm mcu}^B+\varepsilon(t)$, where $r_{\rm mcu}^B$ is a double root $r_{m2}=r_{m3}$ of the potential $\mathcal{V}_{\rm eff}^s(r)$, leading to the solution, $r(t)-r_{\rm mcu}^B \approx \varepsilon (0) e^{\lambda t}$ with the Lyapunov exponent obtained as
\begin{equation}
    \lambda^2 = - {\mathcal{V}_{\rm eff}^s}''(r) \bigg|_{r=r_{\rm mcu}^B}
    = \frac{\left(\gamma_m^2-1\right) (r_{\rm mcu}^B -r_{m1}) (r_{\rm mcu}^B-r_{m4}) \left( {r_{\rm mcu}^B}^2-2 Mr_{\rm mcu}^B+Q^2\right)^2}{\left(\lambda_m M r_{\rm mcu}^B {s_\parallel} -\lambda_m Q^2 {s_\parallel} +\gamma_m {r_{\rm mcu}^B}^4\right)^2}.
\end{equation}
With the parameters along the red line in Fig. \ref{parameter_space} of the double root, $r_{\rm mcu}^B =r_{m3} =r_{m2}$, the Lyapunov exponent is plotted in Fig. \ref{Lyapunov} as a function of $r_{\rm mcu}^B$ from $r_{\rm isco}$ to $r_{\rm mcu,min}^B$ which is the smallest $r_{\rm mcu}^B$ when $\gamma_{\rm mc} \rightarrow 1$.
Since the charge $Q$, which acts as an effective repulsive force effect, for a fixed value of the spin, both $r_{\rm isco}$ and $r_{\rm mcu,min}^B$ shift towards the horizon for a relatively larger charge, giving the larger Lyaponov exponent.
Similarly, for the spin $s_\parallel >0$ ($s_\parallel <0$) with the effect as an effective repulsive (attractive) force, $r_{\rm isco}$ and $r_{\rm mcu,min}^B$ become smaller (larger) than those for spinless particles, giving the larger (smaller) Lyapunov exponent.
The possible violation of the Lyapunov bound \cite{maldacena-2016} will be studied elsewhere \cite{jeong-2023}.

\subsection{Inspiral motion from the ISCO}
\label{sec3B}
We now consider the parameters $(\lambda_{\rm isco}, \gamma_{\rm isco})$ at point D in Fig. \ref{parameter_space} for the ISCO radius, where the particle starts from $r_i \lesssim r_{\text{isco}}$ and moves towards the black hole \cite{mummery-2022, mummery-2023, ko-2024}.
The huge simplification will be seen in the analytical solution in terms of elementary functions.
The inspiral solutions have been constructed in Kerr-Newman black holes for a spinless particle in \cite{ko-2024}, where part of the solutions with the spin effects from the spin-corrected roots $r_{m1}$ in (\ref{roots}), $r_{\rm isco}$ in (\ref{iscoroot}), and particle parameters $\gamma_m =\gamma_{\rm isco}$, $\lambda_m =\lambda_{\rm isco}$ are of the same form for the spinless particle in \cite{ko-2024} but in the limit of $a \to 0$ for RN black holes.

First of all, the integral (\ref{Irintegral}) can be easily carried out with the result $\tau_{m}(r)$ in \cite{mummery-2022, mummery-2023, ko-2024}
and inversion $r^I(\tau_m)$ of (28) with $X^{I}(\tau_m)$ of (29) in \cite{ko-2024} with $r_{\rm isso} \to r_{\rm isco}$ and $\nu_{r_i}=-1$.
The integrals $I_t$, $I_\varphi$, and $I_\psi$ are obtained through (\ref{I_t}), (\ref{I_phi}), and (\ref{I_psi}) with input of the functions $I^I_{\pm}(\tau_m)$, $I^I_{1}(\tau_m)$ and $I^I_{2}(\tau_m)$ given by (36)-(38) in \cite{ko-2024}, and an extra integral of $I^I_{i\pm}(\tau_m)$ expressed as
\begin{equation}
I^I_{i\pm}(\tau_m)=\frac{\sqrt{1-\gamma_m^2}}{r_{\rm isco}-\pm i\lambda_m}\tau_m
+\frac{2
}{\sqrt{(\pm i\lambda_m-r_{m1})\left(r_{\rm isco}-\pm i\lambda_m\right)^3}} \tanh^{-1}\sqrt{\frac{(\pm i\lambda_m-r_{m1})(r_{\rm isco}-r^{I}(\tau_{m}))}{(r_{\rm isco}-\pm i\lambda_m)(r^{I}(\tau_{m})-r_{m1})}} -\mathcal{I}_{\pm_i}^{I}\, ,\label{IIipm}
\end{equation}
which will contribute to the motion of $\delta \vartheta$ in (\ref{I_psi}).
%
%
Therefore, the azimuthal angle $I^I_{\varphi} \sim 1/\sqrt{\epsilon}$ has a power-law divergence near the ISCO radius as in \cite{ko-2024}.
Another limit is near the outer horizon $r \to r_+$, where $I^I_t \sim \log{\epsilon}$ has the logarithmic divergence in the limit, but the azimuthal angle $I^I_{\varphi}$ is finite in \cite{ko-2024}.
It is worth mentioning that the induced motion of $\delta \vartheta$ in (\ref{precession}), or $I^I_{\psi}$, depends on $I^I_r$, $I^I_{i+}$, and $I^I_{i-}$, which are all finite during the journey of the particle. They will remain finite even near the horizon. \\

\begin{figure}[h]
\centering
\includegraphics[width=14cm]{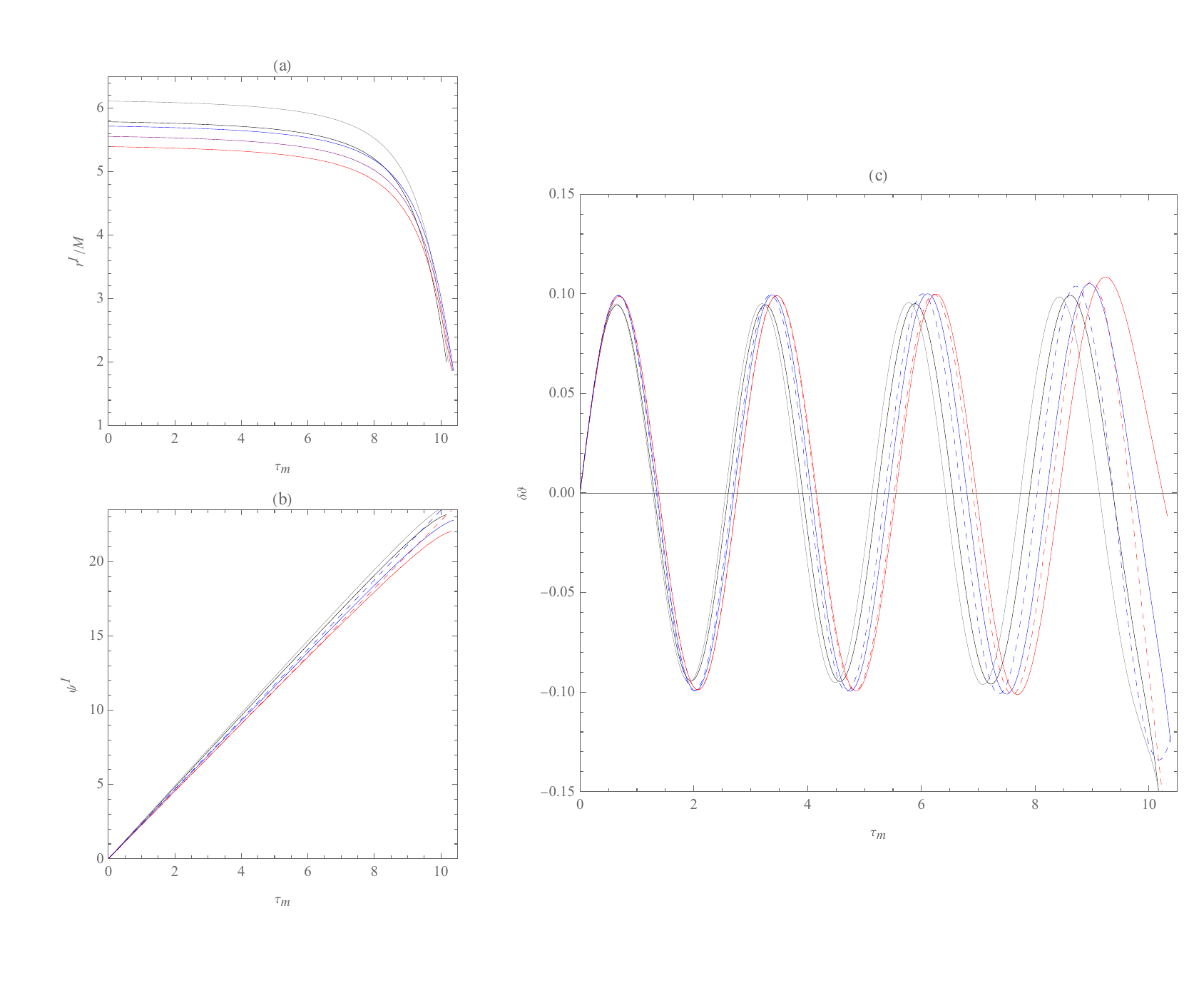}
    \caption{Figures of $r^I(\tau_m)$ in (a), $\psi^I(\tau_m)$ in (b), and $\delta \vartheta(\tau_m)$ in (c) of the inspiral for spinning particle with a misaligned spin for the nonzero $s_\perp$ moving around the RN black hole with the charge $Q$. The same parameters are chosen as in Fig. \ref{oscillation2} but with the different initial condition of $r$ at $r_i=r_{\rm{isco}} -0.05M$ and the choice of $(\lambda_{\rm isco}, \gamma_{\rm isco})$ from the triple root at point D in Fig. \ref{parameter_space}.}
\label{ISCO misaligned}
\end{figure}

\begin{figure}[h]
\centering
\includegraphics[width=14cm]{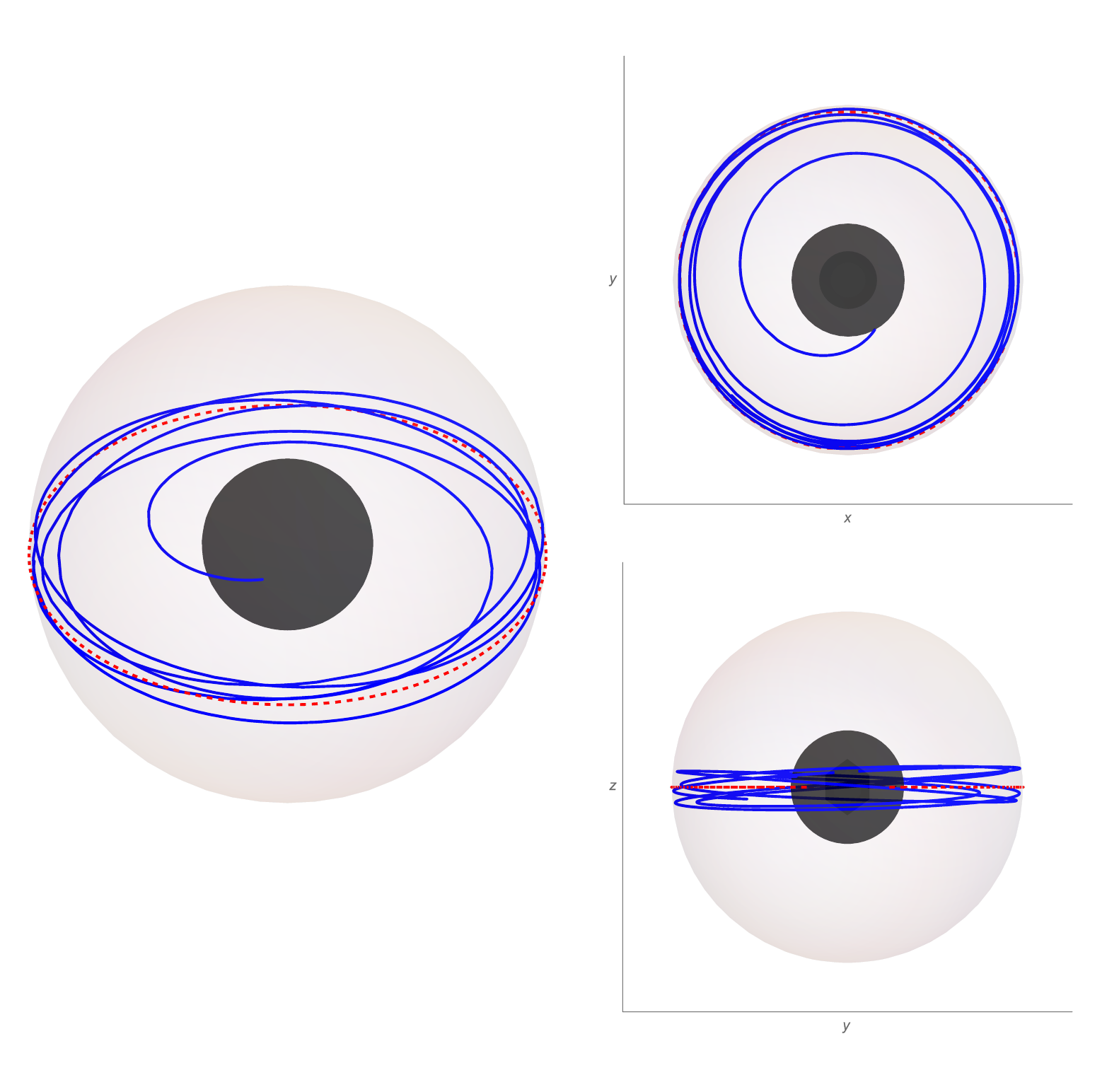}
    \caption{The insprial orbit with the same parameters as in Fig. \ref{oscillation2} but with the different initial condition of $r$ at $r_i=r_{\rm{isco}} -0.05M$ and the ISCO radius (the red dashed line) with $(\lambda_{\rm isco}, \gamma_{\rm isco})$ at point D in Fig. \ref{parameter_space}.}
\label{inspiral}
\end{figure}

For the trajectories with the aligned spin $s_{\parallel}=s$ and $s_\perp=0$, the plot will be similar to that of the spinless case in \cite{ko-2024}, but with the spin-corrected ISCO radius as shown in Fig. \ref{ISCOfig}.
The most noteworthy trajectory is for the misaligned spin with $s_\perp\neq 0$ shown in Fig. \ref{inspiral}.
Apart from the motion in the $r$ and $\varphi$ directions, the oscillation along the $\vartheta$ direction is induced in Fig. \ref{inspiral}.
In particular, the precession angle in (\ref{I_psi}) can be approximated as $\psi^I(\tau_m) \approx \omega_\vartheta^I \tau_m$, where the frequency of oscillation of $\delta \vartheta$ is shown as
\begin{equation}
    \omega_\vartheta^I =
    \gamma_m \lambda_m -\frac{\gamma_m \lambda_m^3}{\lambda_m^2+r_{\rm isco}^2}
    - \gamma_m \lambda_m^2 ~ {\rm Im} \left[ \frac{(r_{\rm isco}-r_i) \sqrt{(r_{\rm isco}+i \lambda_m) (r_{m1}+i \lambda_m)}}{(r_i+i \lambda_m) \sqrt{(r_{\rm isco}+i \lambda_m)^3 (r_{m1}+i \lambda_m)}} \right]\, ,
\end{equation}
which shows a good agreement in the early $\tau_m$ with Fig. \ref{ISCO misaligned} for various choices of $Q$ and $s$.
Thus, the corresponding period in the Mino time will be
\begin{equation}
{\cal{T}}^I_{\vartheta}= 2\pi/\omega_\vartheta^I \,.
\end{equation}
In Fig. \ref{ISCO misaligned}, for a fixed value of the spin, as the charge $Q$ increases, the period increases. For a fixed $Q$, the period increases from $s_\parallel <0$ to $s_\parallel >0$.
However, the period in terms of the coordinate time $t$ of relevance of the asymptotic observer can be obtained through (\ref{t}) and (\ref{I_t}) to be
\begin{equation}
T^I_{\vartheta}= t(\tau_m=2\pi/\omega_\vartheta^I)
\end{equation}
with an input of $I^I_r$, $I^I_\pm$, $I^I_1$, and $I^I_2$.
These solutions may provide useful information for the study of subsequent gravitational wave emission and black hole accretion \cite{mummery-2022, mummery-2023}.

%
\newpage

In \cite{mummery-2023b}, analytical solutions are presented for the thermodynamic properties of thin accretion flows in the region within the ISCO of a black hole.
It is argued that in the case of the adiabatic intra-ISCO accretion, the driving force is mainly due to gravity, where the radial 4-velocity of these inspiral orbits is an important input.
Here we can extend it from (\ref{r_eq}), (\ref{t eq}), and (\ref{phi eq}) with spin corrections but in a RN black hole.
%
In particular, from (\ref{r_eq}) and (\ref{R four root}), we have the radial velocity
\begin{equation} \label{u^r}
\begin{split}
u^r
=& \nu_{r_i} \sqrt{\frac{\left(1-\gamma_m^2\right) (r-r_{m1}^{(0)}) (r_{\rm isco}^{(0)}-r)^3}{r^4}} \\
&+\nu_{r_i} \frac{\left(\gamma_m^2-1\right) (r-r_{\rm isco}^{(0)})^2 \left[3 r_{m1}^{(0)} r_{\rm isco}^{(1)}+r_{m1}^{(1)} r_{\rm isco}^{(0)} -(r_{m1}^{(1)}+3 r_{\rm isco}^{(1)}) r \right]}{2 r^2 \sqrt{\left(\gamma_m^2-1\right) (r-r_{m1}^{(0)}) (r-r_{\rm isco}^{(0)})^3}} s_\parallel
+ \mathcal{O}(s_\parallel^2)
\end{split}
\end{equation}
with $\nu_{r_i}=-1$ and nonzero root {$r_{m1}$} in (\ref{roots}) involving the spin correction in a RN black hole, which can reduce to the form of the equatorial inspiral orbit in \cite{mummery-2022, mummery-2023} for a spinless case in a Kerr black hole with $r_{m1}=0$ and $s_\parallel=0$, given by
\begin{equation}
U^r = -\sqrt{\frac{2M}{3 r_I}} \left( \frac{r_I}{r}-1 \right)^{3/2},
\end{equation}
where the factor ${2M}/{3 r_I}=1-\gamma^2_m$ from (\ref{root_relation}) and $r_I$ is the ISCO radius in the Kerr back hole.

\section{Concluding remarks}
\label{sec4}
In this work we follow and extend the work of \cite{witzany-2024} to consider a spinning particle orbiting a Reissner-Nordstr\"om black hole with a spherically symmetric metric.
The dynamics along its trajectory is governed by the Mathisson-Papapetrou equations in the pole-dipole approximation with the degrees of freedom of a monopolar point mass and its spin with the coupling of the curvature and the spin.
For such a spherically symmetric metric, without loss of generality, we choose the motion in the plane at $\theta=\pi/2$.
With two conserved quantities, namely the energy and angular momentum of the particle, and two approximate conserved quantities to order in linear spin, the magnitude of the spin and its projection along the orbital angular momentum, the equations of motion for the radial, azimuthal angle as well as time components can be derived and can be turned into an integral form in terms of the Mino time for both aligned and misaligned spins with the orbital motion.
In the aligned case, the motion of the particle is restricted to the plane at $\theta={\pi}/{2}$, whereas in the misaligned case, there exists an induced oscillatory motion $\delta \vartheta$ around the plane.
From the equation of motion along the radial direction, one can define the radial potential and find the roots with which to construct the parameter space for various types of orbits.
One of the most interesting types of root is a triple root corresponding to the ISCO.
It is known that the zeroth-order ISCO radius $r^{(0)}_{\rm isco}$ decreases as the charge increases.
The spin correction shows the same behavior.
In addition, the parallel (antiparallel) spin effect with respect to the orbital angular momentum acts as an effective repulsive (attractive) force, making the ISCO radius smaller (larger) than that of spinless particles.

One of the closed-form solutions of the orbit we obtain is an extension of the solution in \cite{witzany-2024} to a Reissner-Nordstr\"om black hole, where the orbit travels between two turning points of the outermost roots with the corrections from the particle spin and the black hole charge.
The corresponding oscillation periods are derived from the analytical solutions in both the Mino time and the coordinate time.
In addition, the oscillation period of the induced $\delta \vartheta$ motion in the early time can also be derived.
The future work is to explore the subsequent gravitational waves following \cite{babak-2007} to consider a family of Kludge waveforms that can be derived from the trajectory of the particle in Boyer-Lindquist coordinates.
The other two orbits in bound motion of interest are the homoclinic orbit and the inspiral motion from near ISCO, and both of them involve the linear spin effect.
Their solutions are simple and can only be expressed involving elementary functions.
Especially, for the misaligned spin, the period of oscillation of the induced $\delta \vartheta$ motion is expressed both in the Mino time and in the coordinate time in terms of such elementary functions.
The homoclinic orbit can shed light on whether the particle orbiting the black hole can behave chaotically \cite{suzuki-1997, zelenka-2020}, where the associated Lyapunov exponent is obtained.
The possible violation of the Lyapunov bound \cite{maldacena-2016, jeong-2023} will be studied in the near future.
These solutions of the spiral and plunge motions into the black hole are also of astrophysical interest due to the fact that they have direct relevance to black hole accretion phenomena.
We also obtain an important quantity to study accretion within the innermost stable orbit on thermodynamic properties in \cite{mummery-2023b}, which is the radial 4-velocity for these inspiral orbits with spin corrections.

\appendix
\section{The roots of the radial potential $\mathcal{R}_m(r)$}
\label{appendixA}
We summarize the roots of the radial potential $\mathcal{R}_m (r)$ for a spinless particle in the following \cite{wang-2022}
with $\mathcal{R}_m (r)$ given by
\begin{align}\label{R_m}
\mathcal{R}_m({r})=S_m {r}^4+T_m {r}^3+U_m {r}^2+V_m {r}+W_m,
\end{align}
where
\begin{align}
&S_m=\gamma_m^2-1,\\
&T_m=2M,\\
&U_m=-\lambda_m^2-Q^2,\\
&V_m=2M\lambda_m^2,\\
&W_m=-Q^2\lambda_m^2.
\end{align}
Furthermore, it is useful to represent the radial potential using its roots, namely
\begin{align}
\mathcal{R}_m({r}) =\left(\gamma_m^2-1\right) \left({r}-r_{m1}^{(0)}\right) \left({r}-r_{m2}^{(0)}\right) \left({r}-r_{m3}^{(0)}\right) \left({r}-r_{m4}^{(0)}\right)\, .
\end{align}
By the Ferrari's method, four roots of this quartic polynomial are listed as
\begin{align}
r_{m1}^{(0)}&=-\frac{M}{2\left(\gamma_m^2-1\right)}+z_m+\sqrt{-\hspace*{1mm}\frac{\textbf{X}_m}{2}-z_m^2-\frac{\textbf{Y}_m}{4z_m}},\\
r_{m2}^{(0)}&=-\frac{M}{2\left(\gamma_m^2-1\right)}+z_m-\sqrt{-\hspace*{1mm}\frac{\textbf{X}_m}{2}-z_m^2-\frac{\textbf{Y}_m}{4z_m}},\\
r_{m3}^{(0)}&=-\frac{M}{2\left(\gamma_m^2-1\right)}-z_m+\sqrt{-\hspace*{1mm}\frac{\textbf{X}_m}{2}-z_m^2+\frac{\textbf{Y}_m}{4z_m}},\\
r_{m4}^{(0)}&=-\frac{M}{2\left(\gamma_m^2-1\right)}-z_m-\sqrt{-\hspace*{1mm}\frac{\textbf{X}_m}{2}-z_m^2+\frac{\textbf{Y}_m}{4z_m}},
\end{align}
where
\begin{align}
z_m=\sqrt{\frac{\Omega_{m +}+\Omega_{m -}-\frac{\textbf{X}_m}{3}}{2}}\ ,
\end{align}
and
\begin{align}
\Omega_{m \pm}=\sqrt[3]{-\hspace*{1mm}\frac{\textbf{Q}_m}{2}\pm\sqrt{\left(\frac{\textbf{P}_m}{3}\right)^3+\left(\frac{\textbf{Q}_m}{2}\right)^2}}\ ,
\end{align}
with
\begin{align}
\textbf{P}_m&=-\hspace*{1mm}\frac{\textbf{X}_m^2}{12}-\textbf{Z}_m \ ,\\
\textbf{Q}_m&=-\hspace*{1mm}\frac{\textbf{X}_m}{3}\left[\left(\frac{\textbf{X}_m}{6}\right)^2-\textbf{Z}_m\right]-\hspace*{1mm}\frac{\textbf{Y}_m^2}{8}.
\end{align}
$X_m$, $Y_m$, and $Z_m$ are the short notation for
\begin{align}
\textbf{X}_m&=\frac{8U_m S_m -3T_m^2}{8S_m^2},\\
\textbf{Y}_m&=\frac{T_m^3-4U_m T_m S_m+8V_m S_m^2}{8S_m^3},\\
\textbf{Z}_m&=\frac{-3T_m^4+256W_m S_m^3-64V_m T_m S_m^2+16U_m T_m^2S_m}{256S_m^4}.
\end{align}
The sum of the roots satisfies the relation
\begin{equation} \label{root_relation}
r_{m1}^{(0)} +r_{m2}^{(0)} +r_{m3}^{(0)} +r_{m4}^{(0)} =-\frac{2M}{\gamma_m^2-1}.
\end{equation}

\section{Analytical Solution for Unbound Motion}
\label{appendixB}
We consider the unbound orbit with the parameters at point E in Fig. \ref{parameter_space}(b) for $\gamma_m>1$.
The solution $r(\tau_m)$ becomes
\begin{align}
{r(\tau_m)}=\frac{r_{m4}(r_{m3}-r_{m1})-r_{m3}(r_{m4}-r_{m1}) \,{\rm sn}^2\left({X^{U}(\tau_m)}\left|{k^{U}}\right)\right.}{(r_{m3}-r_{m1})-(r_{m4}-r_{m1}) \,{\rm sn}^2\left({X^{U}(\tau_m)}\left|{k^{U}}\right)\right.},
\end{align}
where
\begin{align}
{X^{U}(\tau_m)}
=&\frac{\sqrt{(\gamma_m^2-1)(r_{m3}-r_{m1})(r_{m4}-r_{m2})}}{2}\tau_m
+\nu_{r_i} F\Bigg(\sin^{-1}\left(\sqrt{\frac{(r_i-r_{m4})(r_{m3}-r_{m1})}{(r_i-r_{m3})(r_{m4}-r_{m1})}}\right) \left|k^{U}\Bigg)\right.\, ,
\end{align}
with
\begin{align}
k^{U}=&\frac{(r_{m3}-r_{m2})(r_{m4}-r_{m1})}{(r_{m3}-r_{m1})(r_{m4}-r_{m2})} .
\end{align}
For this case, the integrals $I_t$, $I_\varphi$, and $I_\psi$ are expressed as in (\ref{I_t}), (\ref{I_phi}), and (\ref{I_psi}) {but with the factor $\sqrt{\gamma_m^2-1}$ for the unbound motion}, respectively, can be given by
\begin{small}
\begin{align}
I_{\pm}^{U}(\tau_m)&=\frac{2}{\sqrt{(r_{m3}-r_{m1})(r_{m4}-r_{m2})}} \left[{\frac{X^{U}(\tau_m)}{r_{m3}-r_{\pm}}}+\frac{(r_{m3}-r_{m4})\Pi\left({\beta_{\pm}^{U}};\Upsilon_{\tau_m}^{U}\left|{k^{U}}\right.\right) }
{(r_{m3}-r_{\pm})(r_{m4}-r_{\pm})}\right]
-{\mathcal{I}_{\pm_i}^{U}} ,\\
I_{i\pm}^{U}(\tau_m)&=\frac{2}{\sqrt{(r_{m3}-r_{m1})(r_{m4}-r_{m2})}} \left[{\frac{X^{U}(\tau_m)}{r_{m3}\mp i\lambda_m}}+\frac{(r_{m3}-r_{m4})\Pi\left({\beta_{i\pm}^{U}};\Upsilon_{\tau_m}^{U}\left|{k^{U}}\right.\right) }
{(r_{m3}\mp i\lambda_m)(r_{m4}\mp i\lambda_m)}\right]
-{\mathcal{I}_{i\pm_i}^{U}} ,\\
I_{1}^{U}(\tau_m)&=\frac{2}{\sqrt{(r_{m3}-r_{m1})(r_{m4}-r_{m2})}} \left[r_{m3}{X^{U}(\tau_m)}+(r_{m4}-r_{m3}) \Pi\left({\beta^{U}};\Upsilon_{\tau_m}^{U}\left|{k^{U}}\right)\right.\right] -{\mathcal{I}_{1_i}^{U}} ,\\
I_{2}^{U}(\tau_m)&=\nu_{r_i} \frac{\sqrt{\left(r(\tau_m)-r_{m1}\right)\left(r(\tau_m)-r_{m2}\right)\left(r(\tau_m)-r_{m3}\right)\left(r_{m4}-r(\tau_m)\right)}}{r(\tau_m)-r_{m3}} \nonumber\\
&-\frac{r_{m1}\left(r_{m4}-r_{m3}\right)-r_{m3}\left(r_{m4}+r_{m3}\right)}{\sqrt{(r_{m3}-r_{m1})(r_{m4}-r_{m2})}} X^U(\tau_m)-\sqrt{(r_{m3}-r_{m1})(r_{m4}-r_{m2})} E\left(\Upsilon^U_{\tau_m}\left|{k^{U}}\right)\right.\nonumber\\
&+\frac{\left(r_{m4}-r_{m3}\right) \left(r_{m1}+r_{m2}+r_{m3}+r_{m4}\right)}{\sqrt{(r_{m3}-r_{m1})(r_{m4}-r_{m2})}} \Pi\left(\beta^U;\Upsilon^U_{\tau_m}\left|{k^{U}}\right)\right.-{\mathcal{I}^U_{2_i}} ,
\end{align}
\end{small}
and
\begin{align}
&{\Upsilon^U_{\tau_m} = {\rm am}\left(X^U(\tau_m) \mid k^{U}\right) = \nu_{r_i} \sin^{-1} \left( \sqrt{\frac{(r_{m3}-r_{m1})(r(\tau_m)-r_{m4})}{(r_{m4}-r_{m1})(r(\tau_m)-r_{m3})}} \right),} \\
%
&\beta^U_{\pm} = \frac{(r_{m3}-r_{\pm})(r_{m4}-r_{m1})}{(r_{m4}-r_{\pm})(r_{m3}-r_{m1})}, \\
&\beta^U_{i\pm} = \frac{(r_{m3}\mp i\lambda_m)(r_{m4}-r_{m1})}{(r_{m4}\mp i\lambda_m)(r_{m3}-r_{m1})}, \\
&\beta^U = \frac{r_{m4}-r_{m1}}{r_{m3}-r_{m1}}.
\end{align}

\begin{acknowledgments}
This work was supported in part by the National Science and Technology Council (NSTC) of Taiwan, Republic of China.
\end{acknowledgments}

\addcontentsline{toc}{chapter}{Bibliography}
\bibliography{References}

\end{document}